%% file: main.tex
\documentclass[10pt,conference]{IEEEtran}
\usepackage{cite}
\usepackage{amsmath,amssymb,amsfonts}
\usepackage{algorithmic}
\usepackage{graphicx}
\usepackage{textcomp}
\usepackage{xcolor}
\usepackage[hyphens]{url}
\usepackage{mathptmx} 
\usepackage{fancyhdr}
\usepackage[normalem]{ulem}
\usepackage[final]{microtype}
\usepackage[keeplastbox]{flushend}
\usepackage{ragged2e}
\usepackage{comment}
\usepackage[symbol]{footmisc}
\usepackage{subcaption}
\usepackage{paralist}
\usepackage{float}
\usepackage[bookmarks=true,breaklinks=true,letterpaper=true,colorlinks,citecolor=blue,linkcolor=blue,urlcolor=blue]{hyperref}

\def\BibTeX{{\rm B\kern-.05em{\sc i\kern-.025em b}\kern-.08em
    T\kern-.1667em\lower.7ex\hbox{E}\kern-.125emX}}

\newcommand{\scheme}{\textsc{Mercury}}
\newcommand{\scache}{\textsc{MCache}}
\newcommand{\am}[1]{\textcolor{red}{AM: #1}}

\title{\scheme: Accelerating DNN Training By Exploiting Input Similarity
\vspace{-0.1cm}} 

\author{V. Janfaza, K. Weston, M. R. Ghods, S. Mandal, F. Mahmud, A. Hilty, and A. Muzahid\\
Computer Science and Engineering, Texas A\&M University
\vspace{-0.2cm}}

\begin{document}
\maketitle
\thispagestyle{plain}
\pagestyle{plain}

\begin{abstract}
\input{abstract}
\end{abstract}

\section{Introduction}
\label{sec:intro}
\input{intro}
\section{Background}
\label{sec-back}
\input{back}
\section{Main Idea: {\scheme}}
\label{sec:idea}
\input{idea}

\section{Support for different dataflows}
\label{sec:support-other}
\input{support}

\section{Implementation Details}
\label{sec:implement}
\input{implement}

\section{Experimental Setup}
\label{sec:experiment}
\input{experiment}
\section{Evaluation}
\label{sec:result}
\input{result}

\section{Conclusions}
\label{sec:conclusion}
\input{conclusion}


\bibliographystyle{IEEEtranS}
\bibliography{refs}

\end{document}

%% file: abstract.tex
Deep Neural Networks (DNN) are computationally intensive to train. It consists of a large number of multidimensional dot products between many weights and input vectors. 
However, there can be significant similarity among input vectors. 
If one input vector is similar to another, its computations with the weights are similar to those of the other and, therefore, can be skipped by reusing the already-computed results.  
We propose a novel scheme, called \scheme, to exploit input similarity during DNN training in a hardware accelerator. \scheme\ uses 
Random Projection with Quantization (RPQ) to convert an input vector to a bit sequence, called Signature. 
A cache (\scache) stores signatures of recent input vectors along with the computed results. If the Signature of a new input vector matches that of an already existing vector in the \scache, the two vectors are found to have similarities. Therefore, the already-computed result is reused for the new vector. 
To the best of our knowledge, \scheme\ is the {\em first} work that exploits input similarity using RPQ for accelerating DNN training in hardware.
The paper presents a detailed design, workflow, and implementation of the \scheme.
Our experimental evaluation with twelve different deep learning models shows that \scheme\ saves a significant number of computations and speeds up the model training by an average of $1.97\times$ with an accuracy similar to the baseline system.


%% file: intro.tex

Deep Neural Networks (DNNs) have become ubiquitous in recent years.
They are used for diverse tasks such as image and video recognition, recommendation systems, 
natural language processing, etc.~\cite{cnnapp, cnn-rev, face-rec}. Due to the versatility of DNN models, special hardware accelerators have been proposed and built~\cite{tpu, eyeriss, diannao, dadiandao, pudiannao, shidiannao, prime, isaac, fused-cnn}. 
DNNs are computation intensive. For example, Convolutional Neural Network requires
30k to 600k operations per pixel~\cite{eyeriss}. The computation volume is even higher when the accelerator trains a DNN model. However, inputs used during training often have similarities. 
Our objective 
is to improve the computational efficiency of DNN training by exploiting such similarities.

\subsection{Computations with Input Similarity}
DNN operations consist of numerous multidimensional dot products between weight and input vectors extracted from the weight and input matrices. Let us consider a weight vector $\textbf{w}$ and two input vectors $\textbf{v}_1=[v_{1,1}, v_{1,2}, v_{1,3}]$ and $\textbf{v}_2=[v_{1,1}+\epsilon_1, v_{1,2}+\epsilon_2, v_{1,3}+\epsilon_3]$. If $\epsilon_i$ (for $1\le i\le 3$) represents an insignificant difference, then $\textbf{v}_1$ and $\textbf{v}_2$ have value similarity. The dot product of $\textbf{v}_2$ and $\textbf{w}$ would be
$\textbf{v}_2\textbf{.w}=\textbf{v}_1\textbf{.w}+{\bf \epsilon.w}$. If $\epsilon_i\approx0$, then ${\bf\epsilon.w}\approx0$, and therefore, $\textbf{v}_2\textbf{.w}\approx\textbf{v}_1\textbf{.w}$. In other words, if $\textbf{v}_2$ and $\textbf{v}_1$ have value similarity,
the computation of $\textbf{v}_2$ with a weight vector is considerably similar to that of $\textbf{v}_1$ and, therefore, can be skipped by reusing the results of 
$\textbf{v}_1$. 

\begin{figure}[h]
\centering
\begin{subfigure}{0.4\columnwidth}
\centering
 \vspace{-0.35cm}
\includegraphics[width=\columnwidth]{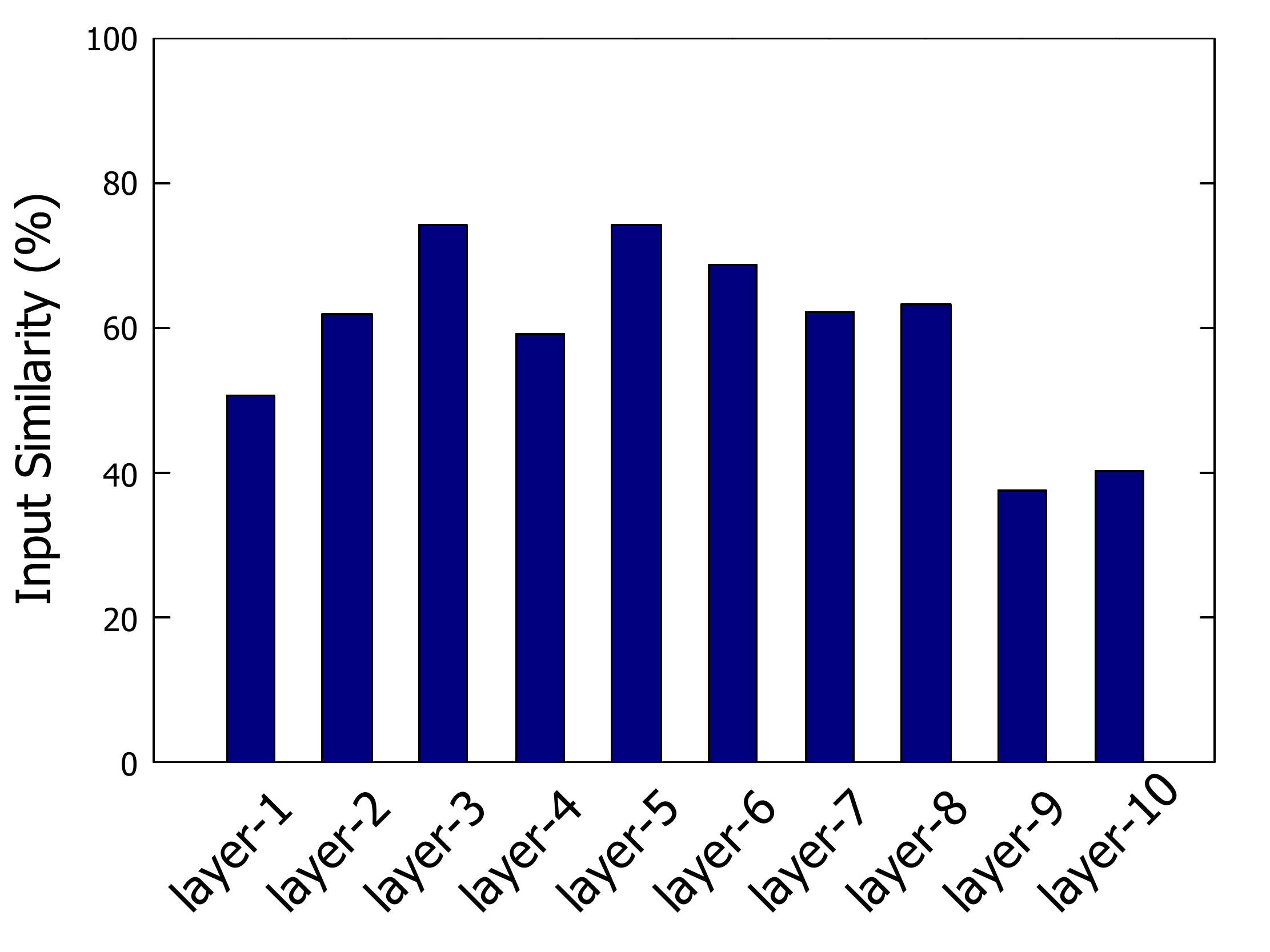}
 \vspace{-0.6cm}
 \caption{Input vector.}
\label{case-input-sim}
\end{subfigure}
\begin{subfigure}{0.4\columnwidth}
\centering
 \vspace{-0.35cm}
\includegraphics[width=\columnwidth]{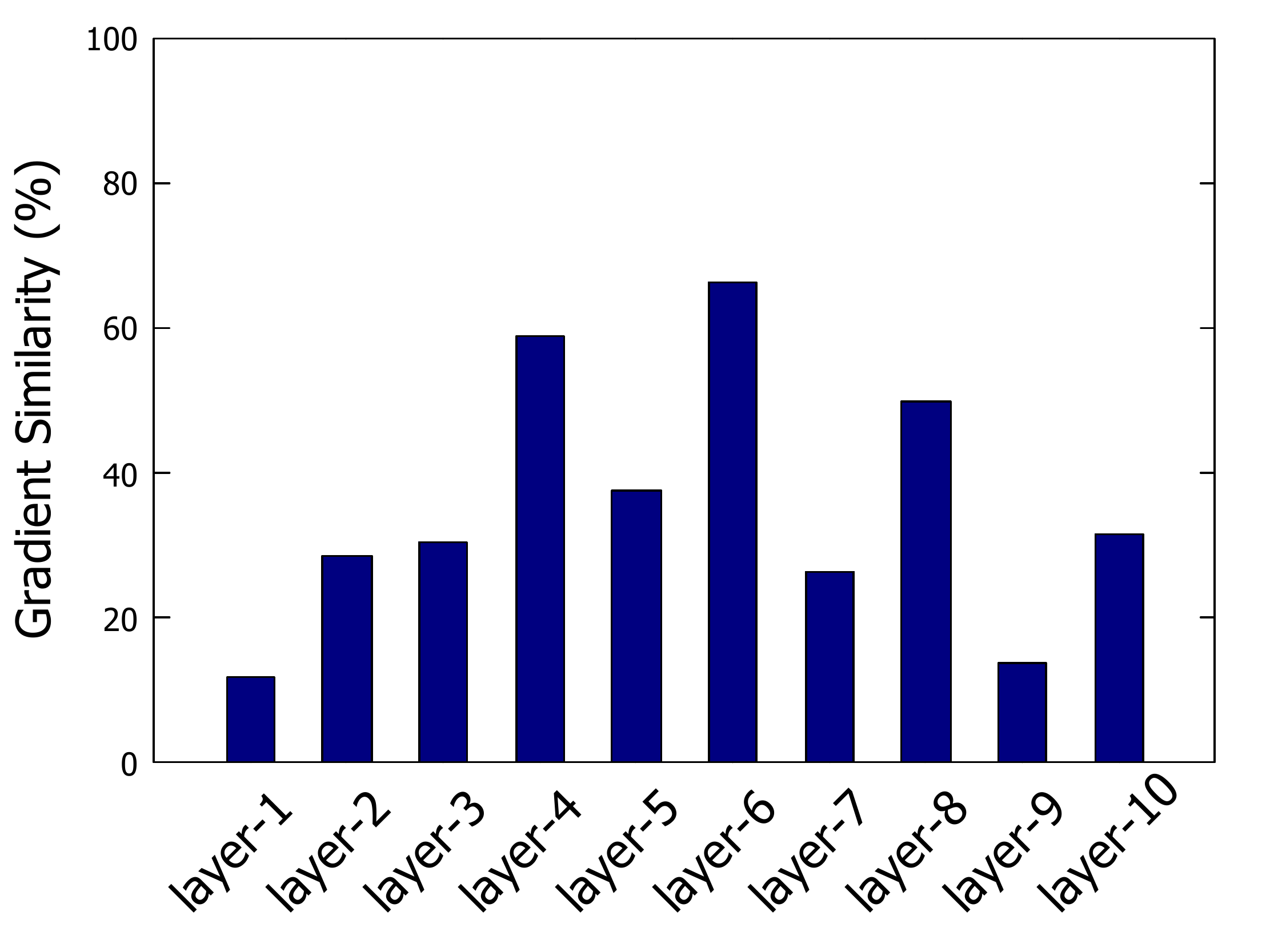}
 \vspace{-0.6cm}
\caption{Gradient vector}
\label{case-gradient-sim}
\end{subfigure}
\vspace{-0.05cm}
\caption{Similarity among input and gradient vectors of VGG13.}
 \vspace{-0.4cm}
\label{sim-in-grad}
\end{figure}

To further motivate the readers, we analyzed the VGG13 network~\cite{vgg} with ten convolution layers. We counted what fraction of input and gradient vectors have similarities in the convolution layers. The similarity is detected using a well-established technique called Random Projection with Quantization (RPQ)~\cite{random-projection} (more details in $\S$~\ref{sec-rpq}). The similarity in input vectors leads to computation reuse in the forward propagation, while that of gradient vectors leads to computation reuse in the backward propagation. Figure~\ref{sim-in-grad} shows that
VGG13 has up to $75\%$ similarity among input vectors and 
up to $67\%$ similarity among gradient vectors. By capitalizing on these similarities, \scheme\ speeds up the VGG13 training by $1.89\times$ compared to baseline ($\S$~\ref{ovr-analysis}).

\subsection{State of the Art}
\label{sec-limit} 


When it comes to DNN inference acceleration, the two dominant techniques are sparsity exploitation~\cite{cnvlutin, scnn, snapea, uspe, eyeriss, bit-tactical, sparten} and computational reuse~\cite{ucnn, summerge, dcompress, eie, ttq, diffy, riera-reuse}. Unfortunately, applying these techniques directly during training has never been easy~\cite{tdash, dsg}. Thus, most of the efforts for reducing DNN training time focus on alternative approaches, such as distributed training~\cite{dist-large, dist-survey}, data compression~\cite{echo, gist, jpeg-act}, low precision training~\cite{binarynet, das-mixed, flexpoint}. Recently, there is some work about exploiting sparsity in DNN training~\cite{tdash, dsg, procrustes}. Yet the challenges of reusing similar computational data due to input similarity during training are not well-studied. Most of the current work in this category is either software based~\cite{adaptivedeepreuse, dreuse} or limited to 
inference only~\cite{dreuse, cicek21, diffy, riera-reuse}. Extending them for training in an accelerator is difficult due to two major challenges. 

\begin{compactitem}
    \item {\em Similarity Detection:} Detecting similarity among inputs requires extra computations and hardware. Therefore, reducing the computations while reusing the existing hardware as much as possible becomes a major bottleneck for exploiting input similarity.
    
    \item {\em Dataflow Modification:} When two inputs are similar, computations for one input can be reused for the other. This creates an irregularity in the dataflow of an accelerator. Changing the dataflow or using a new one will vanquish the benefit of the accelerator's original dataflow. Thus, addressing the irregularity in computations while maintaining the original accelerator dataflow becomes a significant objective for adopting input similarity.
\end{compactitem}

\subsection{Proposed Approach}
\label{sec-proposed-app}

We propose a novel scheme, called \scheme, to exploit input similarity during the training phase in a DNN accelerator. 
\scheme\ uses RPQ~\cite{random-projection}
in hardware to detect similarity among input vectors.
We show a formulation of RPQ where it follows the same computation pattern as a convolution operation. Therefore, \scheme\ reuses the existing hardware Processing Elements (PEs) to perform RPQ. \scheme\ uses RPQ to convert an input vector into a bit-sequence, called {\em Signature}. 
\scheme\ calculates one signature for each input vector. 
If two input vectors produce the same signature, they are significantly similar and thus, have higher similarity.
During the DNN operation between a weight and input vector, the input vector's signature is used to access a special cache, called \scache. \scache\ uses signatures to calculate indices and tags and (previously) computed results as data. If there is a hit on \scache, the computation is skipped. Instead, the computed result stored in the data-portion of the cache entry is reused. On the other hand, if there is a miss, the computation continues, and the result is stored into \scache. Input similarity introduces irregularity in the original computation pattern of a DNN accelerator by skipping some computations. \scheme\ adds a bitmap (called {\em Hitmap}) and some shared structures to make the dataflow and computations continuous and uninterrupted. The signatures produced during the forward propagation are stored in memory to be reused during the backward propagation of the training phase. Moreover, \scheme\ dynamically decides when and to what extent input similarity should be exploited based on its impact on performance and accuracy. 
In summary, we make the following contributions:
\begin{compactenum}

    \item \scheme\ is the {\em first} accelerator to exploit input similarity using RPQ for improving training performance.
    We propose to adapt \scheme\ dynamically based on accuracy and performance impact.

    \item We propose to use RPQ in hardware to detect similarity among input vectors dynamically. We show a novel formulation of RPQ where it follows the same computation pattern as a convolution operation. Therefore, \scheme\ can calculate RPQ-based signatures using the same hardware PEs and dataflow used for DNN operations. We show how signature calculation can be further pipelined. 
    
    \item Input similarity causes irregularity in the original computation pattern of an accelerator due to the reuse of computations. We propose to add a cache, \scache, along with a bitmap ({\em Hitmap}) and some shared structures to make the dataflow and computations continuous and regular.

    
    

    \item We implemented \scheme\ in Virtex 7 FPGA board~\cite{virtex}. We showed a scalable implementation of \scache\ to meet the demand of the \scheme. We evaluated \scheme\ using twelve DNN models (including a transformer model) with three different dataflows and achieved an average speedup of $1.97\times$ with an accuracy similar to the baseline system.
    
\end{compactenum}

%% file: back.tex
\subsection{Random Projection with Quantization (RPQ)}
\label{sec-rpq}

\begin{figure}[htpb]
\centering
 \vspace{-0.45cm}
\includegraphics[width=\columnwidth]{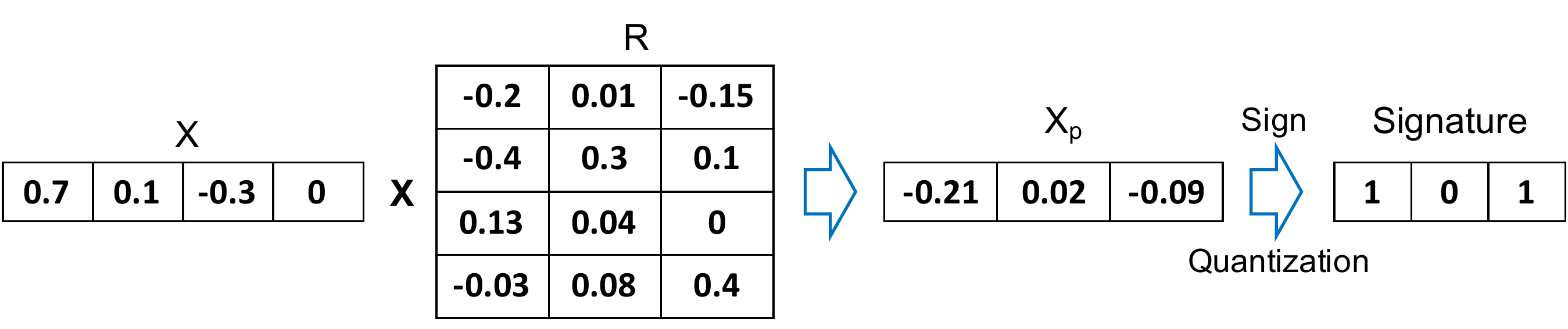}
\vspace{-0.5cm}
\caption{An example of how RPQ converts a vector $\mathbf{X}$ into a projected vector $\mathbf{X}_p$ and eventually, a signature.}
\label{fig-signature-cal}
\vspace{-0.25cm}
\end{figure}

Random Projection~\cite{random-projection} is a dimensionality reduction technique often used in similarity estimation of high-dimensional data (such as image and text). Given a vector, $\mathbf{X}$ of size $1\times m$, random projection works by multiplying $\mathbf{X}$ with a random matrix $\mathbf{R}$ of size $m\times n$. The elements of $\mathbf{R}$ are randomly populated
from a normal distribution, whose mean is 0 and variance is 1. The multiplication produces a projected vector $\mathbf{X}_p$ of size $1\times n$. Thus, random projection converts one vector 
to another with a different dimension (often a lower one). 
Random projection ensures that if two vectors are close (similar) in their original dimension, their projected vectors will also be close (with a Euclidean distance scaled accordingly) in the newer dimension. Elements of $\mathbf{X}_p$ can be quantized further. One such quantization approach is sign-based. So, if an element of $\mathbf{X}_p$ has a sign bit equal to 0, it is quantized to 0. Otherwise, it is quantized to 1. Thus, RPQ converts $\mathbf{X}$ into a bit sequence, called signature. Figure~\ref{fig-signature-cal} shows an example of how RPQ converts a vector into a signature. If RPQ converts two vectors, $\mathbf{X}_1$ and $\mathbf{X}_2$, into the same signature, their Euclidean distance in the new dimension is 0. Therefore, their distance in the original dimension is $\approx 0$. So, $\mathbf{X}_1\approx\mathbf{X}_2$.


\begin{figure}[h]
\centering
\begin{subfigure}{0.4\columnwidth}
\centering
 \vspace{-0.3cm}
\includegraphics[width=\columnwidth]{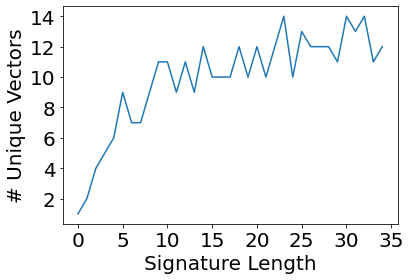}
 \vspace{-0.5cm}
\caption{RPQ}
\label{case-input-sim}
\end{subfigure}
\begin{subfigure}{0.4\columnwidth}
\centering
 \vspace{-0.3cm}
\includegraphics[width=\columnwidth]{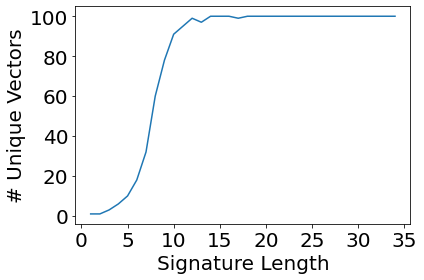}
 \vspace{-0.5cm}
 \caption{Bloom Filter}
\label{case-gradient-sim}
\end{subfigure}
\vspace{-0.2cm}
\caption{Unique vectors found by a) RPQ b) Bloom Filter.}
\label{fig-rpq-func}
\vspace{-0.3cm}
\end{figure}

RPQ has been used in many domains such as learning~\cite{manyfold-rpq}, compression~\cite{rounding-rpq}, etc. To provide insight into how RPQ behaves, we conducted an experiment with ten randomly generated unique vectors of dimension $10$. We generated ten more similar vectors from each of the vectors (by adding some random $\epsilon$ to each dimension). 
We generate signatures of all vectors and compare them with each other to determine how many unique vectors we can find. Since we started with ten unique vectors, a comparison should find a similar number of unique vectors.
Figure~\ref{fig-rpq-func} shows the number of unique vectors found by RPQ. It also shows results with another technique, {\em Bloom Filter}~\cite{bloom1970space}, ~\cite{ceze2006bulk}. For smaller signatures, both methods declare many dissimilar vectors as similar. However, RPQ is able to detect unique vectors better than Bloom Filters at longer signatures. 

\subsection{DNN Accelerator and Dataflow}
\label{sec-baseline}
\begin{figure}[htpb]
\centering
\vspace{-0.3cm}
\includegraphics[width=0.6\columnwidth]{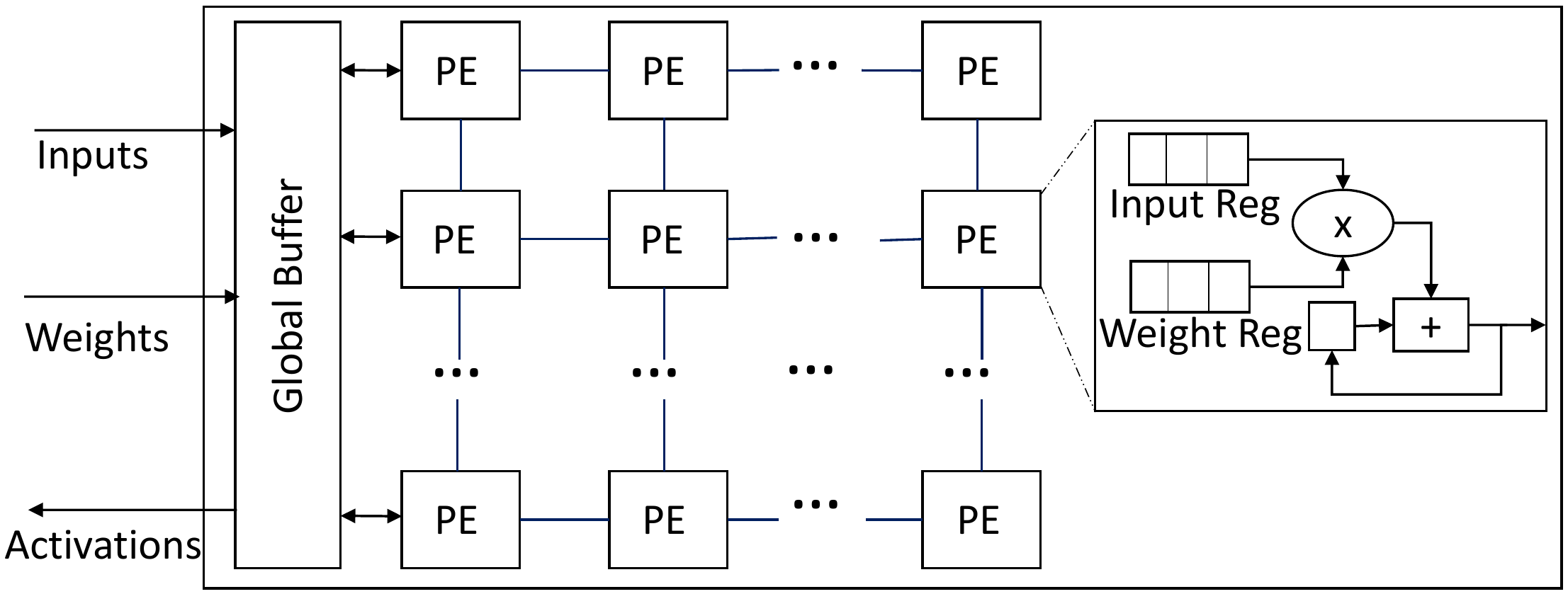}
\caption{Baseline hardware accelerator.}
\label{fig-base}
\vspace{-0.3cm}
\end{figure}

A typical DNN accelerator is shown in Figure~\ref{fig-base}. The accelerator has a number of hardware PEs. 
Each PE has vertical and horizontal connections with neighboring PEs using on-chip networks. There is a global buffer to hold inputs, weights, and partial-sums. The chip is connected to off-chip memory to receive inputs and store outputs. Each PE contains registers to hold inputs, weights, and partial sums. Each PE also contains multiplier and adder units.
Each PE distributes inputs and weights and generates partial sums based on a dataflow.

Different dataflows have been proposed in literature~\cite{dataflows, eyeriss, maeri} to optimize different aspects of the DNN operations. Examples are Weight-Stationary, Output-Stationary, Input-Stationary, and Row-Stationary. 
The dataflow name often reflects which data is kept unchanged in the PE unit throughout the computation.
In Weight-Stationary, each PE statically holds a weight inside its register file. Those operations that use the same weight are mapped to the same PE unit~\cite{dataflows}. 
Output-Stationary~\cite{outputstationary} localizes the partial result accumulation inside each PE unit. 
For Row-Stationary, each PE processes one row of the input. Filter weights stream horizontally, input rows stream diagonally, and partial sums are accumulated vertically. Row-Stationary has been proposed in Eyeriss~\cite{eyeriss} and is considered one of the most efficient dataflows for reuse.

\vspace{-0.1cm}
\subsection{{Extending Row-Stationary for Training}}
\label{sec-back-RS}
\begin{figure}[!htpb]
\centering
\vspace{-0.5cm}
\includegraphics[width=0.8\columnwidth]{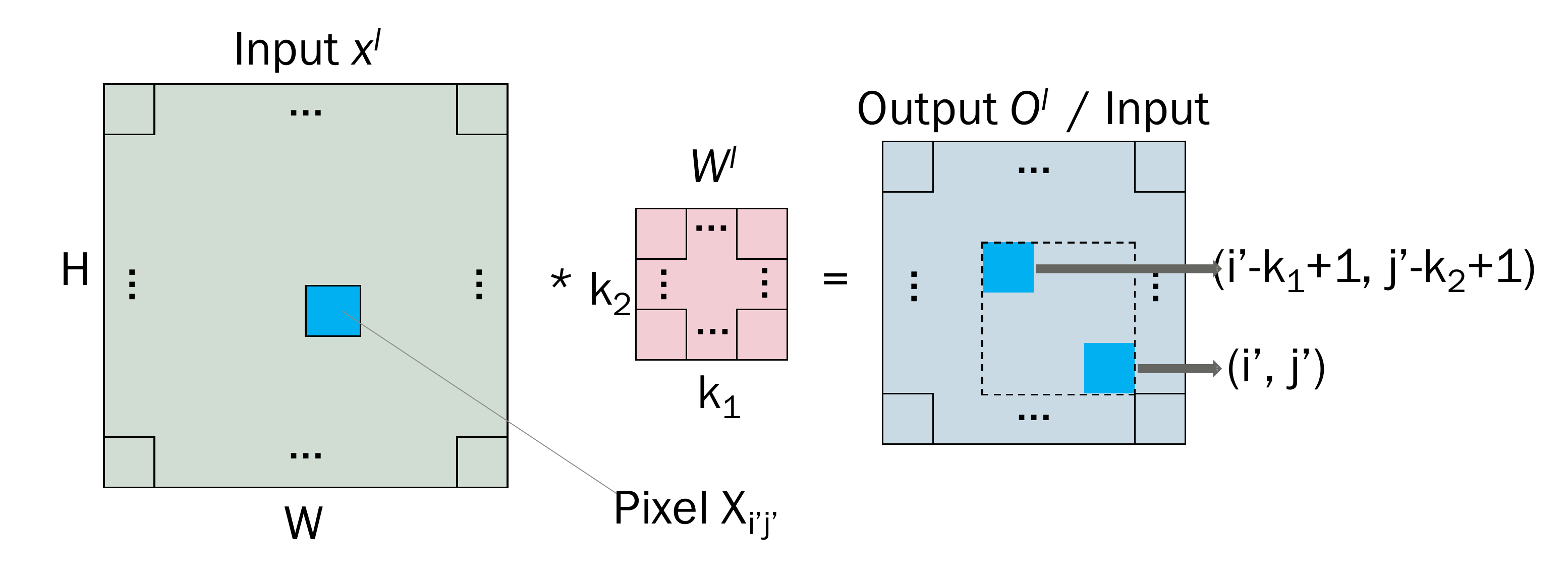}
\vspace{-0.2cm}
\caption{The region in the output affected by the input $x_{i',j'}$.}
\label{fig-RS-BP-Input}
\vspace{-0.5cm}
\end{figure}

Eyeriss~\cite{eyeriss} used row stationary dataflow for inference. However, it can be easily extended for training. During the back propagation, each layer performs two computations - one for the weight update and the other for calculating the gradient of the inputs. Consider a convolution between the input of dimension $H\times W$ and the weight of dimension $k_1\times k_2$. This results in an output of size $(H-k_1+1)\times (W-k_2+1)$. For simplicity, let us assume that there is one channel in the input and output. We can easily update the equations with more channels. 
For updating weights, we measure $\frac{\partial{E}}{{\partial{w_{m',n'}^l}}}$ which is interpreted as how changing a single-point $w_{m',n'}$ of the weight affects the loss function $E$. 
\vspace{-0.1cm}
\begin{equation} \label{eq:weight}
\frac{\partial{E}}{{\partial{w_{m',n'}^l}}}=\sum_{i=0}^{H-k_1}\sum_{j=0}^{W-k_2} \delta_{i,j}^l O_{i+m',j+n'}^{l-1}
\vspace{-0.1cm}
\end{equation}
In this equation~\ref{eq:weight}, $\delta_{i,j}^l$, and $O_{i+m',j+n'}^{l-1}$ represent the gradients of outputs in layer $l$ and outputs of layer $l-1$ respectively. Equation~\ref{eq:weight} shows that for calculating the gradient of weights in layer $l$, the convolution between gradients of outputs in layer $l$ and outputs of layer $l-1$ is needed. Similar to inference, this convolution can be done with row-stationary dataflow. 
For the second computation of the back propagation, we compute $\frac{\partial{E}}{{\partial{x_{i',j'}^l}}}$, which can be interpreted as how changing in a single-
point $x_{i',j'}$ of the input feature map affects the loss function $E$.
As shown in Figure~\ref{fig-RS-BP-Input}, we can see that the output region affected by point $x_{i',j'}$ of the input is the output region bounded by the dashed lines where the top left corner point is given by $(i'-k_1+1,j'-k_2+1)$ and the bottom right corner point is given by $(i',j')$. Based on this figure, 
the gradient of input $x_{i',j'}$ can be calculated as:
\vspace{-0.1cm}
\begin{equation} \label{eq:gradient}
\frac{\partial{E}}{{\partial{x_{i',j'}^l}}}=\sum_{m=0}^{k_1-1}\sum_{n=0}^{k_2-1} \underbrace{\delta_{i'-m,j'-n}^{l+1} w_{m,n}^{l+1}}_\text{Convolution}\underbrace{f'\left(x_{i',j'}^l\right)}_\text{Partial Derivative}
\vspace{-0.1cm}
\end{equation}

The first part of this equation \ref{eq:gradient} contains the convolution operation between gradients of layer $l+1$ and weights of layer $l+1$. This is $\sum_{m=0}^{k_1-1}\sum_{n=0}^{k_2-1} \delta_{i'-m,j'-n}^{l+1} w_{m,n}^{l+1}$, and the second part is the partial derivative of the activation function. 
So, similar to inference, we can also use the row-stationary dataflow for this part of back propagation. We can pad the gradient matrix by zeros for pixels in corners to have a generic formula for all pixels. Thus, we can use the row-stationary dataflow for training as it contains similar computations as the inference. 

\subsection{Computation Reuse}
\label{sec-comp-reuse}


UCNN~\cite{ucnn} exploits weight repetitions in a CNN model. 
At the core of UCNN is the factorized dot product dataflow and activation group reuse. 
SumMerge extends the idea of UCNN into CPU-based implementation~\cite{summerge}.
TensorDash~\cite{tdash} accelerates DNN training by skipping ineffective multiply-accumulate (MAC) computation.
Eager-Pruning~\cite{eager-prune} speeds up training
by detecting and pruning insignificant weights early.

DeepReuse~\cite{dreuse} and Adaptive Deep Reuse (ADR)~\cite{adaptivedeepreuse} exploit similarity in inputs to improve inference and training performance, respectively. The scope of both approaches is limited to software implemented CNN models.
Both approaches use Locality Sensitive Hashing (LSH) to find the similarity among input vectors. Although \scheme\ shares some similarities with ADR at a high level, ADR cannot be directly implemented in hardware because of the following issues. {\em First}, the use of LSH requires a computationally expensive pre-processing step to build clusters and determine the centroids. This step is not trivial to do in a hardware accelerator. That is why Cicek et al.~\cite{cicek-reuse} proposed to design a separate accelerator just for detecting input similarities and interface it with a CNN accelerator to improve the inference performance. {\em Second}, ADR interleaves LSH calculations with convolution operations to determine if some result is already computed or not. Such interleaving of operations in an accelerator will interfere with its original dataflow, thereby reducing the potential benefit of computational similarity. Thanks to our formulation of RPQ as a convolution operation ($\S$~\ref{sec-sig-cal}), additional calculations for detecting input similarities blend in with the original operations without interfering with the accelerator's dataflow. Furthermore, due to the use of Hitmap along with some extra hardware resources, the reuse of already computed results does not modify the dataflow either.

Diffy~\cite{diffy} and Riera et al.~\cite{riera-reuse} propose to exploit similarity in pixelated or streaming video
to accelerate DNN inference. Both approaches use element-wise comparison to detect similar elements. This is inefficient since the comparison either needs to be done serially over all elements or only within a limited number of neighboring elements. Recently, Servais et al.~\cite{jason21} exploit similarity in consecutive rows or columns of inputs and weights for CNN training. Unlike prior approaches, \scheme\ uses a well established hashing technique, RPQ, to check similarity at vector granularity across all vectors in constant time. RPQ is hardware-friendly and can be easily used in both forward and backward propagation. Moreover, unlike \scheme, prior approaches require a new dataflow in the accelerator and cannot be used on top of an existing dataflow. Another line of work looks at redundancy in training data. However, \scheme\ is orthogonal to that approach and can be applied on top of it~\cite{bengio2007scaling},~\cite{coleman2019selection},~\cite{Feldman2020},~\cite{killamsetty2021grad},~\cite{killamsetty2021glister},~\cite{kumar2010self},~\cite{mirzasoleiman2020coresets},~\cite{toneva2018empirical},~\cite{Fan2017Learning}.
\vspace{-0.05cm}

%% file: idea.tex


\vspace{-0.1cm}
\subsection{Overview}
\label{sec-overview}
\begin{figure}[htpb]
\centering
\includegraphics[width=0.75\columnwidth]{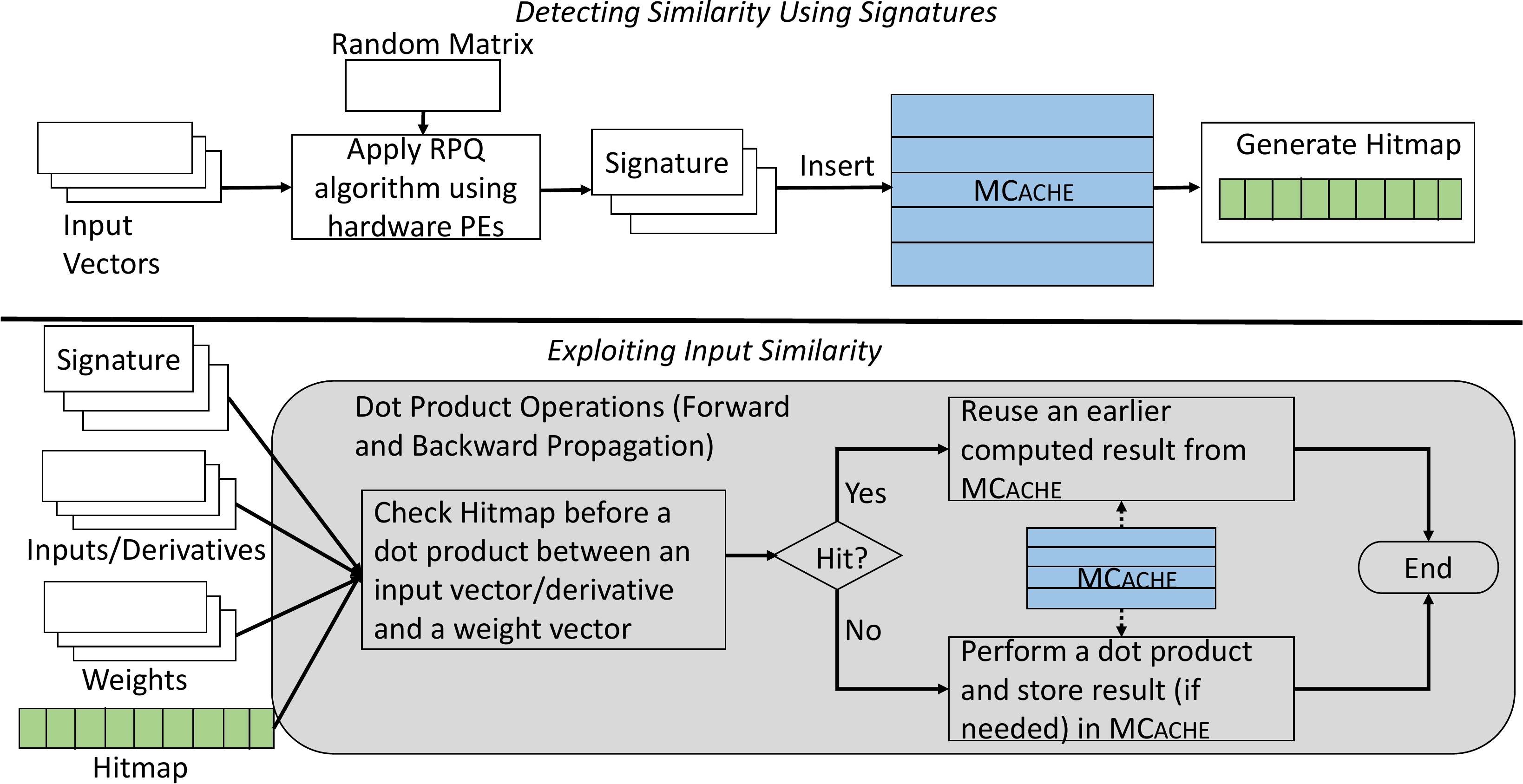}
\vspace{-0.15cm}
\caption{ Overview of \scheme\ operations. \scheme\ first generates RPQ signatures and Hitmap for detecting similarity among input vectors. The signatures are used during dot products to skip computations and reuse results.}
\label{fig-overview}
\vspace{-0.7cm}
\end{figure}

\scheme\ works on vectors extracted from multidimensional inputs. The overview of \scheme\ is shown in Figure~\ref{fig-overview}. 
\textit{First}, \scheme\ determines which input vectors are similar.
This is done before the input vectors are multiplied with weight vectors.
\scheme\ calculates an RPQ-based signature for an input vector by multiplying the vector with a random projection matrix followed by quantization. \scheme\ formulates signature calculation as a convolution operation and reuses the same hardware PEs with the original dataflow.
\scheme\ generates one signature for each input vector.
If two signatures are identical, the corresponding vectors are similar. Therefore, the computed dot product between one of the input vectors and a weight vector can be reused for that of the other input vector and the same weight vector. \scheme\ uses a cache, called {\scache}, to store computed results corresponding to different signatures.
A Hitmap keeps track of the signatures that cause a hit. 
When \scheme\ performs a dot-product between an input vector and a weight vector (or derivatives during the backward propagation), \scheme\ checks if this vector is similar to any prior vector using the Hitmap; if so, the result stored in \scache\ is reused. \scheme\ stores the signatures calculated during the forward propagation and reuses them during the backward propagation to skip similar computations. 
As the training proceeds, \scheme\ increases signature length to adjust to the extent of similarity among vectors. Only vectors with a higher degree of the similarity are allowed to reuse computed results in the later stages of training. Thus, \scheme\ dynamically adjusts computation reuse to keep accuracy degradation insignificant.

Next we will elaborate on \scheme\ operations. We choose row-stationary~\cite{eyeriss} dataflow as our baseline ($\S$\ref{sec-back-RS} shows how it can support training). We will first explain 
how \scheme\ works in this baseline accelerator. Then, we will discuss how other dataflows are supported.

\subsection{Detecting Similarity Using Signatures}
\label{sec-detect-local}
\scheme\ detects similarity dynamically among input vectors before performing any operation with weights. To simplify hardware and keep dataflow the same, \scheme\ formulates signature calculation as a convolution operation, where the convolution is performed between the input vectors and some random projection matrix. This is done every time there is a new set of input vectors (e.g., when a new channel is processed). 
When the actual filters of a channel are convoluted with the input vectors, signatures are used to determine if similar dot products are already computed.


\subsubsection{Signature as a Convolution Operation}
\label{sec-sig-cal}
\begin{figure}[!htpb]
\centering
\begin{subfigure}{\columnwidth}
    \centering
    \includegraphics[width=0.6\columnwidth]{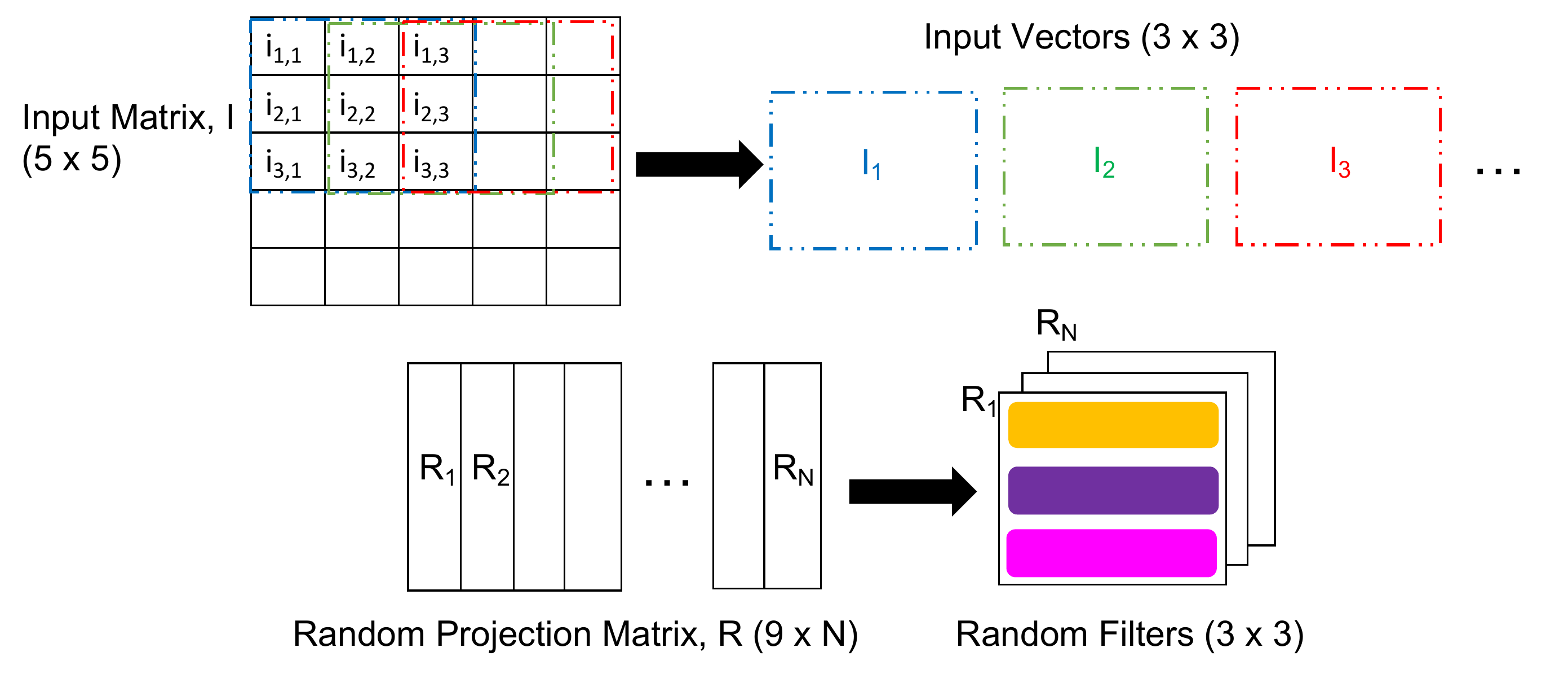}
    \vspace{-0.15cm}
    \caption{Organization of input matrix and projection matrix}
    \label{fig-organization-vector}
\end{subfigure}

\begin{subfigure}{\columnwidth}
    \centering
    \includegraphics[width=0.6\columnwidth]{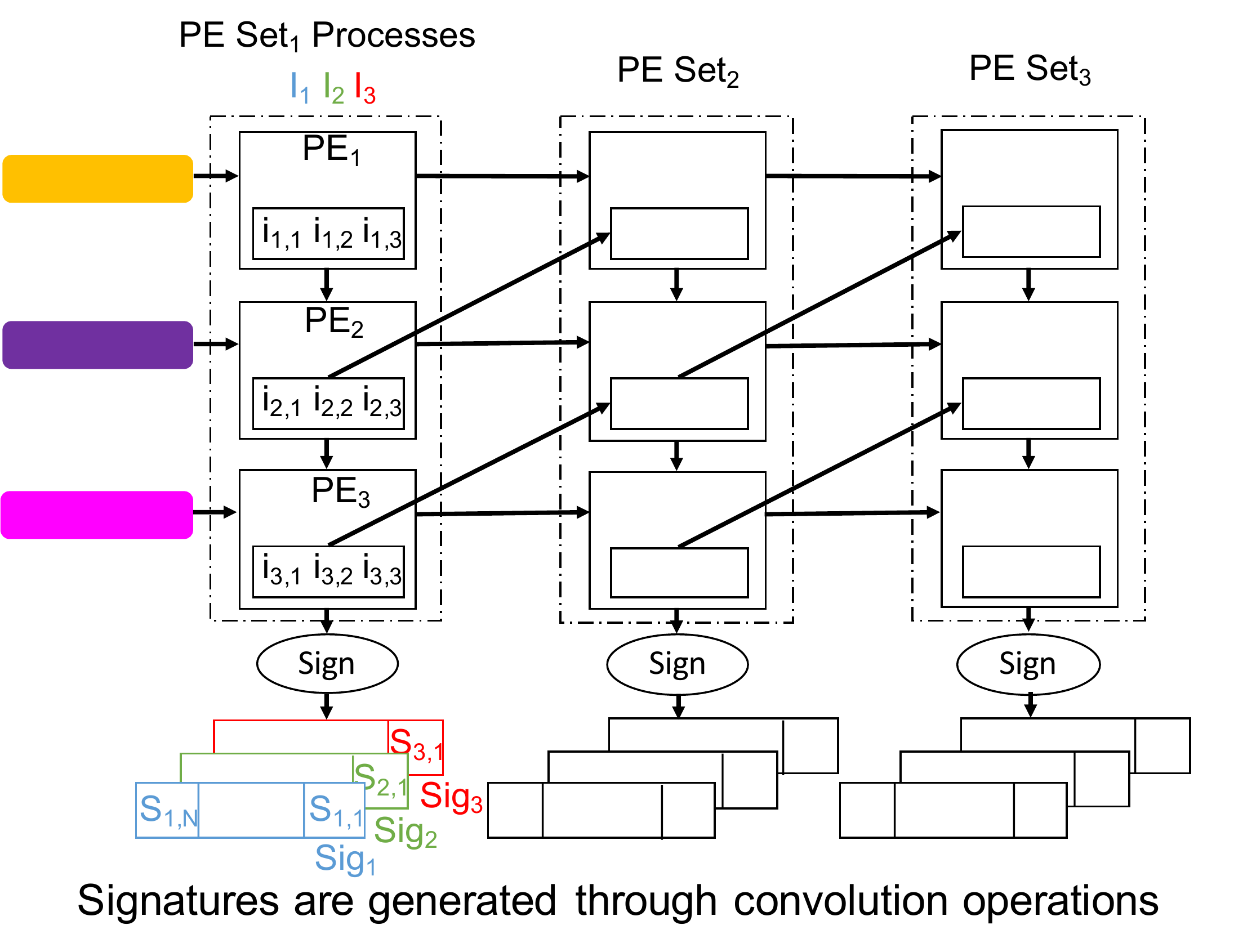}
    \vspace{-0.15cm}
    \caption{Signature calculation}
    \label{fig-signatures}
\end{subfigure}

\vspace{-0.1cm}
\caption{Overview of how \scheme\ generates signatures for input vectors. Input $I_i$ is associated with $Sig_i$ for $i>0$. We refer to the set of PEs performing a 2D convolution as a PE Set. Each PE Set calculates 3 signatures in this example.}
\vspace{-0.75cm}
\end{figure}

Let us assume that 
a $5\times5$ input is convoluted with $3\times 3$ kernels in each channel. Since the kernel size is $3\times 3$, the input vectors extracted from the input matrix are of size $3\times 3$ as shown in Figure~\ref{fig-organization-vector}. Since each input vector has $9$ elements, the random projection matrix, $\mathbf{R}$, is of size $9\times N$ for $N>0$. We can re-organize each column vector of $\mathbf{R}$ 
into the shape $3\times 3$. With this organization, we can treat each column vector of $\mathbf{R}$ (i.e., $\mathbf{R_1}$ to $\mathbf{R_N}$) as a random filter of size $3\times 3$.

Let us consider an input vector $\mathbf{I_1}$. $R$ converts $\mathbf{I_1}$ into a signature $\mathbf{Sig_1}$ consisting of $N$ bits (i.e., $S_{1,1}$ to $S_{1,N}$) using RPQ algorithm ($\S$\ref{sec-rpq}). $S_{1,1}$ is calculated by performing a dot product between $\mathbf{R_1}$ and $\mathbf{I_1}$ followed by a sign comparison. Other bits, $S_{1,2}$ to $S_{1,N}$, are calculated similarly. Similarly, $\mathbf{Sig_2}$ can be calculated from $\mathbf{I_2}$ and so on. In other words, if we perform a 2D convolution followed by a sign comparison by sliding $\mathbf{R_1}$ over all input vectors from $\mathbf{I}$, we can calculate the first bit of each signature (i.e., $S_{1,1}$, $S_{2,1}$, $S_{3,1}$, $\ldots$). Similarly, the $2D$ convolution between $\mathbf{R_2}$ and the input vectors will produce the second bit of each signature and so on. In other words, the signature calculation can be formulated as $2D$ convolutions between input vectors and random filters $\mathbf{R_1}$ to $\mathbf{R_N}$. Therefore, we can easily map it to a row-stationary accelerator.

Assume that the accelerator has $3\time3$ PEs. 
Figure~\ref{fig-signatures} shows the signature calculation process. \scheme\ starts with the random filter $\mathbf{R_1}$ and computes $2D$ convolutions with each input vector. 
Filter rows stream horizontally while input rows stream diagonally. So $PE_1$ to $PE_3$ perform 3(three) 2D convolutions in a streaming fashion - $I_1\bullet R_1$, $I_2\bullet R_1$, and then, $I_3\bullet R_1$. $\bullet$ indicates a 2D convolution. We refer to the set of PEs working on a 2D convolution as a PE Set. Thus, $PE~Set_1$ consists of $PE_1$ to $PE_3$. With $\mathbf{R_1}$ streaming first, $PE~Set_1$ calculates $S_{1,1}$, $S_{2,1}$ and then, $S_{3,1}$. 
After that, $\mathbf{R_2}$ is loaded, and the second bit of each signature is calculated. Thus, $N$ bits of all signatures are calculated through $2D$ convolutions with random filters.

\subsubsection{Dataflow of Signature Calculation}
\label{sec-dataflow-sig}
\begin{figure*}[!htpb]
\centering
\begin{subfigure}[t]{0.25\textwidth}
    \centering
    \includegraphics[width=\columnwidth]{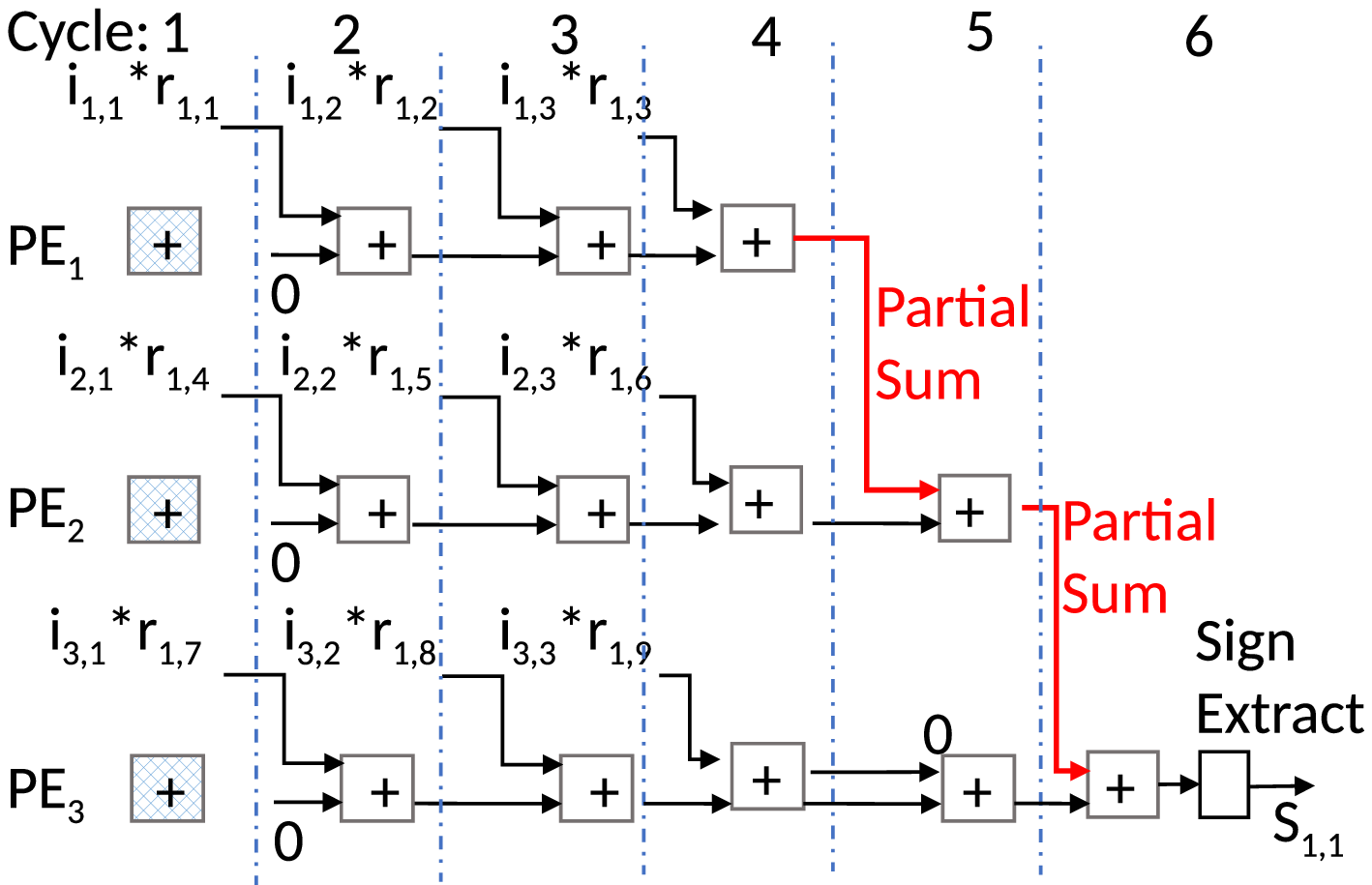}
    \caption{Row-stationary}
    \label{fig-orig-signatures}
\end{subfigure}
\begin{subfigure}[t]{0.50\textwidth}
    \centering
    \includegraphics[width=\columnwidth]{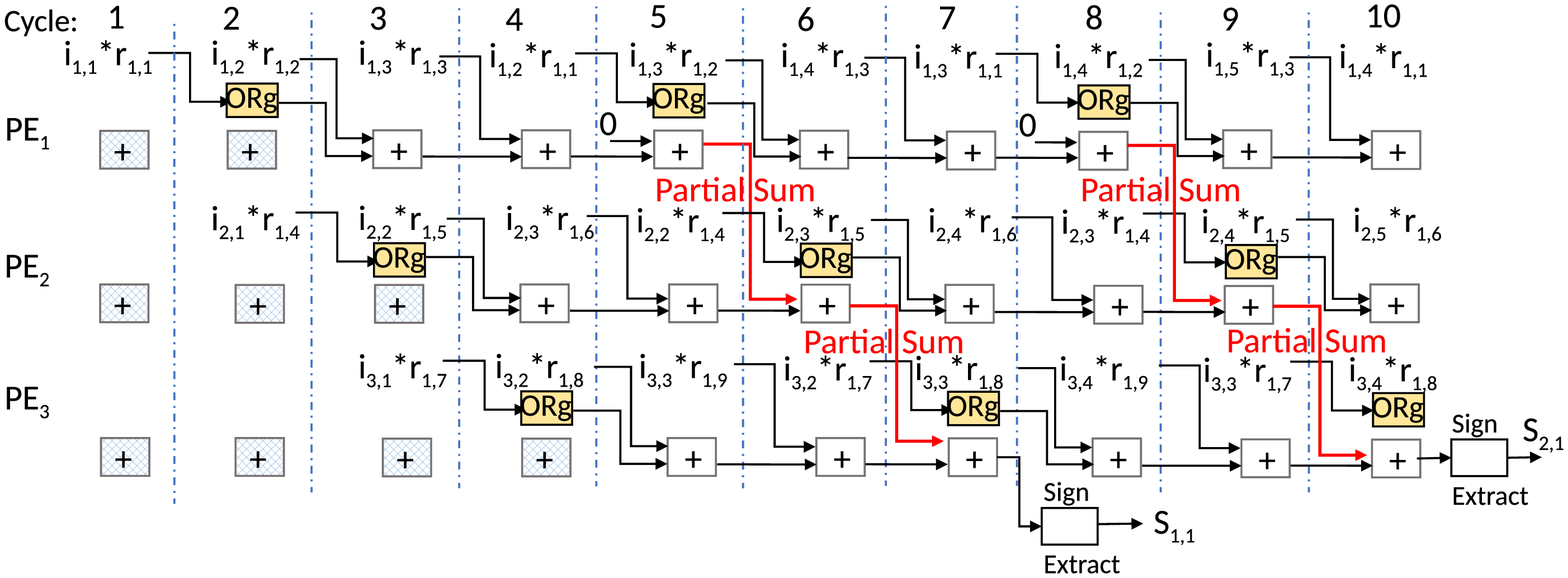}
    \caption{With pipelining}
    \label{fig-nonstop}
\end{subfigure}
\begin{subfigure}[t]{0.22\textwidth}
    \centering
    \includegraphics[width=\columnwidth]{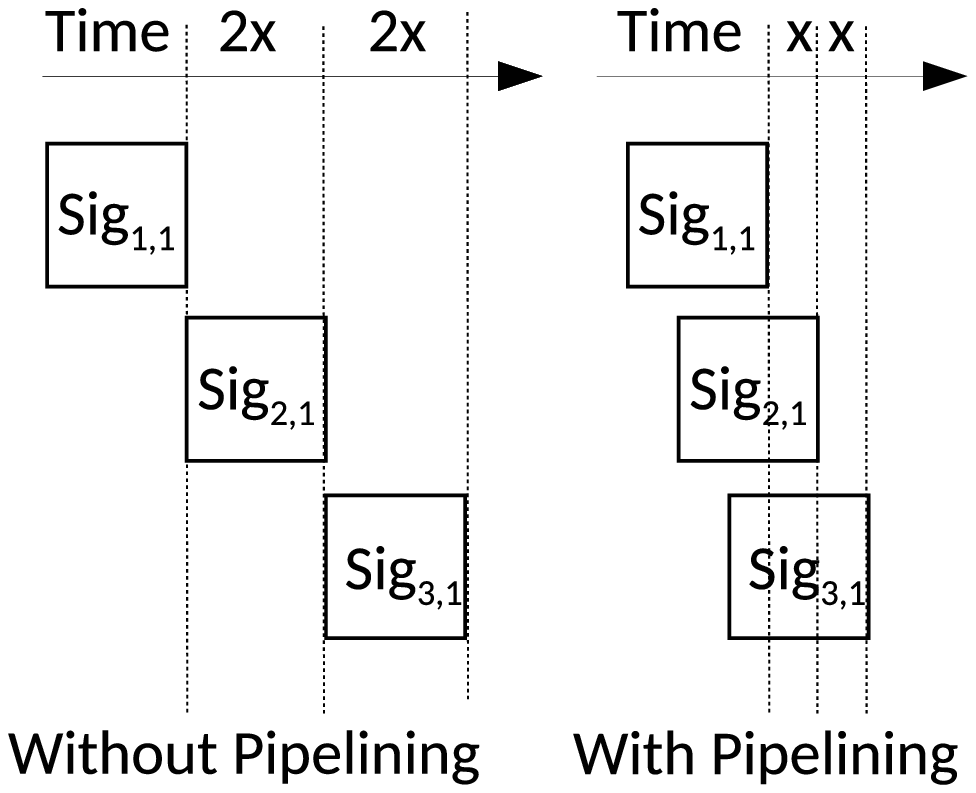}
    \caption{Speed up}
    \label{fig-difference}
\end{subfigure}

\caption{(a) Timing of how \scheme\ generates signatures for input vectors in a row-stationary machine. Red arrow indicates a partial sum of a row and a shaded adder means an idle adder. (b) shows the same with pipelining. Here $Sig_{1,1}$ calculation spans from cycle 1 to 7 while that of $Sig_{2,1}$ spans from cycle 4 to 10. Thus, signature calculations are overlapped. (c) Speed up with pipelined signature calculations. }
\vspace{-0.4cm}
\label{fig-ovr-perf}
\end{figure*}

Let us continue with our example in $\S$\ref{sec-sig-cal}. $i_{1,1}$ to $i_{3,1}$ are the elements of $\mathbf{I_1}$ and $r_{1,1}$ to $r_{1,9}$ are the elements of $\mathbf{R_1}$. Figure~\ref{fig-orig-signatures} shows the timing of signature calculation in a row-stationary accelerator. With $3\times3$ input vectors, it takes four cycles to multiply and accumulate the result of each row and two cycles to accumulate across rows. Thus, it takes six cycles to generate a single bit of a signature. Similarly, the first bit of subsequent signatures takes six cycles each. In general, for $x\times x$ input vectors, it takes $2x$ cycles to calculate each bit of a signature. However, the calculation of one bit of one signature does not overlap with that of another signature (as shown in Figure~\ref{fig-difference}(a)).
Note that we do not assume a separate multiplier and adder unit in each PE. Instead, if we assume a multiply-accumulate (MAC) unit, it takes $2x-1$ cycles to calculate a single bit of a signature (because row accumulation takes one less cycle). However, still, each signature is calculated in a non-overlapping fashion.

We propose to pipeline the calculation of one signature with another. 
The core idea is to add a register, named Overlapped register (ORg), in each PE and intentionally delay the calculation starting time of $PE_2$ and $PE_3$ by 1 and 2 cycles, respectively (as shown in Figure~\ref{fig-nonstop}). This is reminiscent of software pipelining~\cite{soft-pipe}. ORg register is used to hold the result of multiplying the first element of each row of input and random vectors. For example, ORg of $PE_1$ is used to hold $i_{1,1}*r_{1,1}$ in cycle 2 and $i_{1,2}*r_{1,1}$ in cycle 5. When the register holds $i_{1,2}*r_{1,1}$ in cycle 5, it frees up the adder unit which can be used to pass the row accumulation result from $PE_1$ to $PE_2$ in cycle 6. Similarly in cycle 6, ORg of $PE_2$ holds $i_{2,2}*r_{1,4}$ which frees up the adder. Therefore, the adder accumulates the result from $PE_1$ with $PE_2$ and passes it to $PE_3$ in cycle 7.  
In cycle 7, ORg of $PE_3$ holds $i_{3,2}*r_{1,7}$ which frees up the adder to finish the accumulation of all rows. Thus, $Sig_{1,1}$ takes seven cycles to calculate. However, the calculation of $Sig_{2,1}$ has already started. Therefore, following the same flow, $Sig_{2,1}$ will finish in cycle 10, i.e., it takes only three more cycles. Similarly, the first bit of subsequent signatures produced by the same PEs will take 3 more cycles each. Generally, for $x\times x$ input vectors, the first bit of the first signature calculated by a set of PEs takes $2x+1$ cycles, while other bits of any signature take $x$ cycles to finish. This is illustrated in Figure~\ref{fig-difference}(b). 



\subsubsection{Signature Management}
\label{sec-storing-sig}
\scheme\ manages signatures using three structures - Signature Table, \scache, and Hitmap. The Signature Table stores the signatures, \scache\ keeps dot product results computed between different input and weight vectors, and the Hitmap keeps track of which signature causes a hit in \scache.
The Signature Table is indexed by the input vector number so that \scheme\ can easily find it for a particular input vector. 
When a signature is calculated by the PEs, \scheme\ stores it in the signature table and then accesses \scache. 
\scache\ is indexed and tagged with the signature. \scache\ keeps computed dot product results so that input vectors with similar signature can reuse them. The data portion of \scache\ contains the results. \scache\ differs from a normal cache in two ways. {\em First,} since a tag (i.e., signature) is produced before the data (i.e., computed results), cache tag and data are not updated together. To accommodate that, each cache line has two valid bits - Valid Tag (VT) and Valid Data (VD). When a signature is used to initialize the tag section of a cache line, its VT is set while VD remains unset.
The data portion of \scache\ is populated and used as a weight vector (or derivative) is multiplied with the input vector. Different weight vectors populate the data portion with different results (details in $\S$~\ref{sec-exploit}).
{\em Second,} there is no replacement in \scache. When a set is full, no new entries are inserted into \scache. We choose this no replacement policy approach to simplify the design of \scache.

\begin{figure}[htpb]
\centering
\vspace{-0.25cm}
\includegraphics[width=0.45\columnwidth]{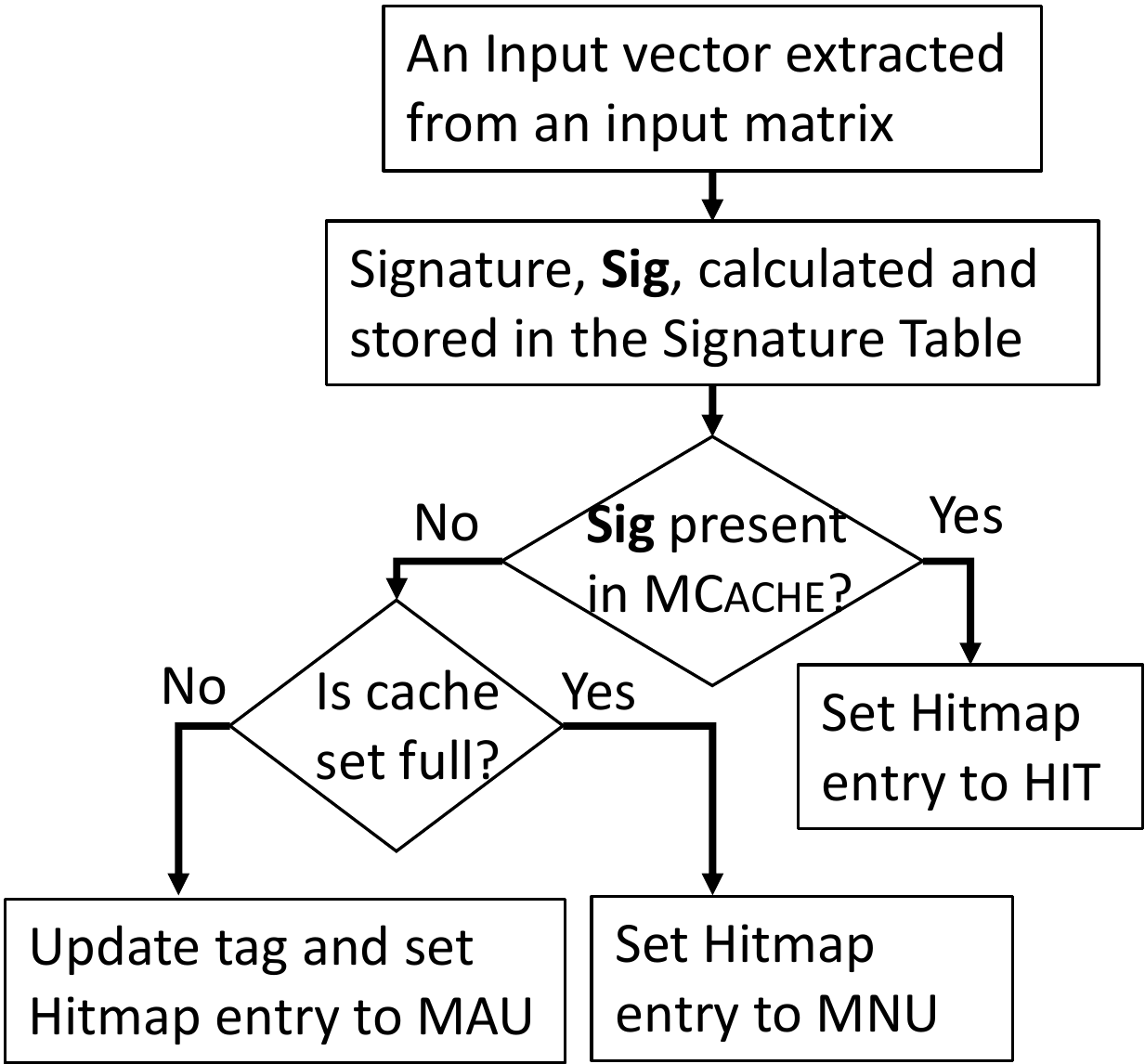}
\vspace{-0.1cm}
\caption{How \scache\ is updated with signatures.\vspace{-0.2cm}}
\vspace{-0.15cm}
\label{fig-update}

\end{figure}

Figure~\ref{fig-update} shows what happens when a signature, $Sig$ for a new input vector is calculated. If $Sig$ is already in \scache, we have a hit. So, the Hitmap entry is set to HIT. Otherwise, $Sig$ is a new signature; therefore, \scache\ checks if the corresponding set is full. If it is not full, $Sig$ is inserted into the cache by updating the tag portion of an entry. Since the data portion will be updated later, the Hitmap entry is marked as Miss And Update (MAU). If the set is full, $Sig$ will not be inserted into \scache. Since no cache entry will be updated, the Hitmap entry is marked as Miss No Update (MNU). When a new set of input vectors are extracted from an input matrix, \scache, Signature table, and Hitmap are cleared.

\subsection{Exploiting Input Similarity}
\label{sec-exploit}
Here, we explain how signatures along with \scache\ can be utilized to skip similar computations in the forward propagation followed by the backward propagation.
We will first explain it for a convolution layer ($\S$~\ref{sec-forward} \& \ref{sec-backward}). In $\S$~\ref{sec-fc-clustering} and$\S$~\ref{sec-attention}, we will address it for other layers. 

\subsubsection{Computation Reuse in Forward Propagation}
\label{sec-forward}

\begin{figure}[htpb]
\centering
\vspace{-0.3cm}
\includegraphics[width=0.6\columnwidth]{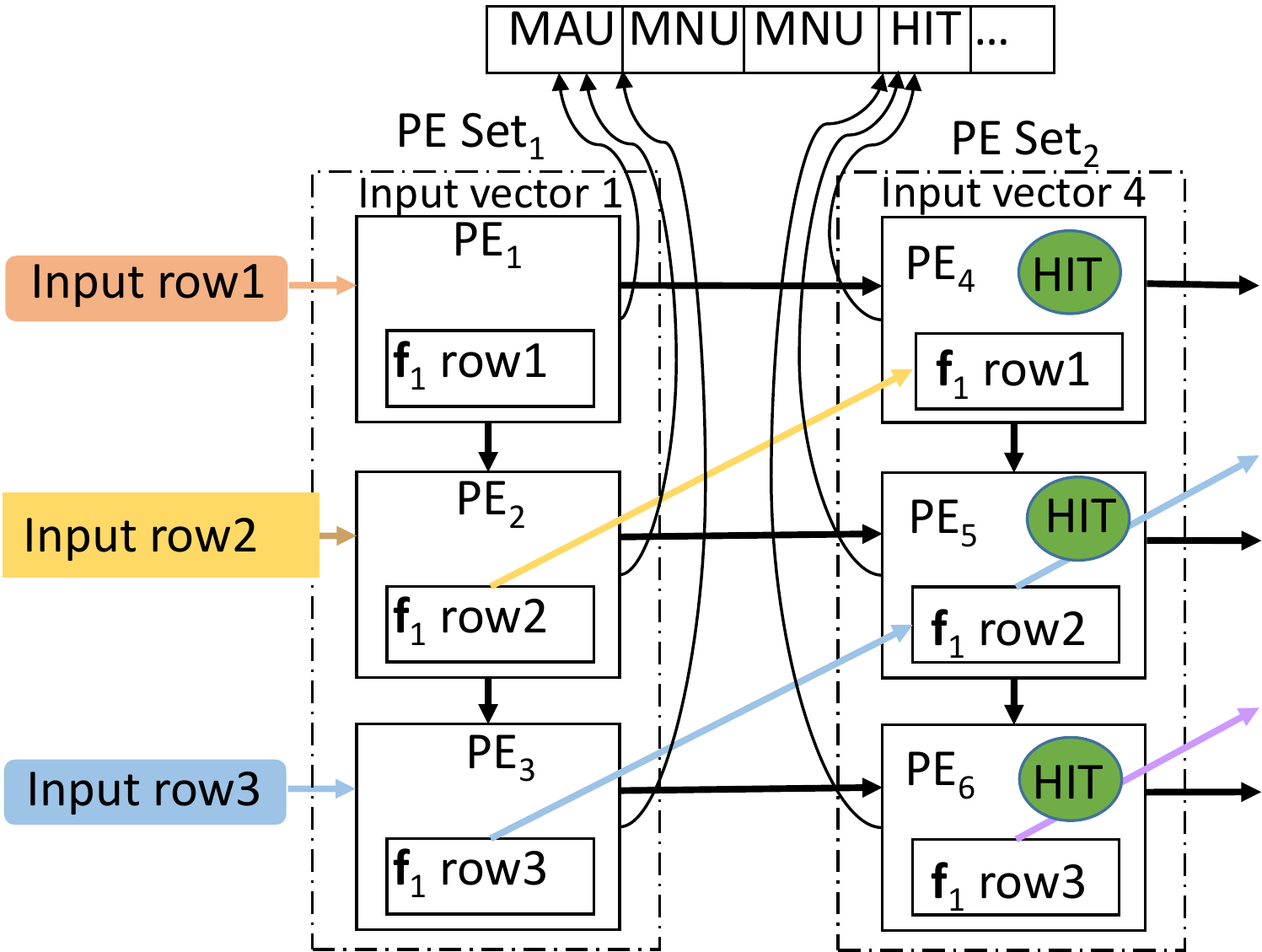}
\vspace{-0.1cm}
\caption{Overview of how computation is reused in the forward propagation. $f_1$ indicates a particular filter.}
\label{fig-forward}
\vspace{-0.35cm}
\end{figure}

During the forward propagation of a convolution layer, 
a $2D$ input convolves with a number of filters in a channel. The $2D$ input consists of a number of input vectors, each with the same size as a filter. Thus, for each channel, an accelerator performs a number of dot products between the input vectors and filters in that channel.
The PEs load one filter and a number of input vectors at a time and perform the dot products. 
The input vectors and filters are passed through the PEs in a streaming fashion. Rows of input vectors are passed through diagonally, whereas rows of filters are passed through horizontally. Partial sums are accumulated vertically. Figure~\ref{fig-forward} shows this flow. Here, we assume that $PE_1$, $PE_2$, and $PE_3$ perform a dot product between an input vector and a filter. Thus, {\em PE Set 1} consists of $PE_1$ to $PE_3$. Similarly, {\em PE Set 2} consists of $PE_4$ to $PE_6$.

Each PE has an input buffer, a number of input and weight registers, a multiplier, and an adder. 
The input buffer holds inputs as they arrive at the PE. Input registers load values from the input buffer. As filter weights arrive, they are stored in the weight registers. The multiplier and adder work with the values from the input and weight registers. Each PE in PE set starts with the first input vector from its input buffer. Each PE checks whether the corresponding entry in the Hitmap is a HIT. 
Remember that the Hitmap and signatures are calculated before the convolution operations for a channel begin.
A HIT indicates that the input vector is similar to an earlier one; hence, the dot product result stored in the \scache\ can be reused instead of calculating again. 
That is why the PE skips the dot product. Instead, using the signature of the input vector, the stored result is fetched from \scache\  and used as the PE set result. On the other hand, if the entry in the Hitmap is MAU or MNU, the PEs in the PE set perform the dot product. 
If the Hitmap entry is MAU, the corresponding \scache\ line contains the signature of this input vector, but the result (data) portion is still empty. So, the data portion is updated with the dot product result, and its VD is set.
If the Hitmap entry is MNU, \scache\ does not contain the signature at all. So, \scheme\ does not store the dot product result in the cache. The PEs in the PE set proceed with the next input vector from the respective input buffer. Note that although a PE set might skip computations occasionally, the dataflow (filter streaming horizontally and inputs streaming diagonally) remains unchanged due to the Hitmap.

If we assume the same inputs as in $\S$~\ref{sec-sig-cal}, 
then $PE~Set_1$ operates on input vectors $1$, $2$, and $3$ whereas $PE~Set_2$ operates on input vectors $4$, $5$, and $6$ in Figure~\ref{fig-forward}. Therefore, at first, the PEs in $PE~Set_1$ check entry $1$ of the hitmap. Based on that entry, the PEs either reuse results from \scache\ or compute dot products using input vector $1$ and the current filter. For the example in Figure~\ref{fig-forward}, $PE~Set_1$ will compute dot product because entry $1$ is MAU. On the other hand, the PEs in $PE~Set_2$ check entry $4$ in the hitmap and act accordingly. Note that each PE set acts independent of other PE sets. As a result, one PE set might reuse a lot of computed results and finish computations early whereas another PE set may lag behind computing many dot products. We propose two designs to address this - synchronous and asynchronous.

{\bf - Synchronous Design: }
In synchronous design, when a PE set finishes computation, it waits until all other PE sets are done. Each PE set maintains a busy bit (B). A controller checks all B bits. If none of them are busy (i.e., $B=0$), the controller instructs the PEs to load with the next filter and input vectors.
At this point, \scache\ may contain computed results from the previous filter and input vectors. Those results cannot be used because weights change in the new filter. Therefore, \scache\ invalidates all VD bits. A bitline connecting all VD bits is used for this purpose. The input vectors remain the same within the same channel. Therefore, VT flags and Hitmap are still valid and kept as they are. When \scheme\ proceeds with the next channel, the signatures in the signature table, \scache, and Hitmap are recalculated and reinitialized.

{\bf - Asynchronous Design: }
The synchronous design is intuitive and simpler but limits performance improvement because the faster PE sets remain idle until the slowest one completes. Therefore, we propose an asynchronous design where the faster PE sets can work on the next filter and input vectors while the slower ones work on the previous filter and input vectors. However, this requires additional buffer and co-ordination scheme.
\begin{figure}[htpb]
\centering
\vspace{-0.3cm}
\includegraphics[width=0.6\columnwidth]{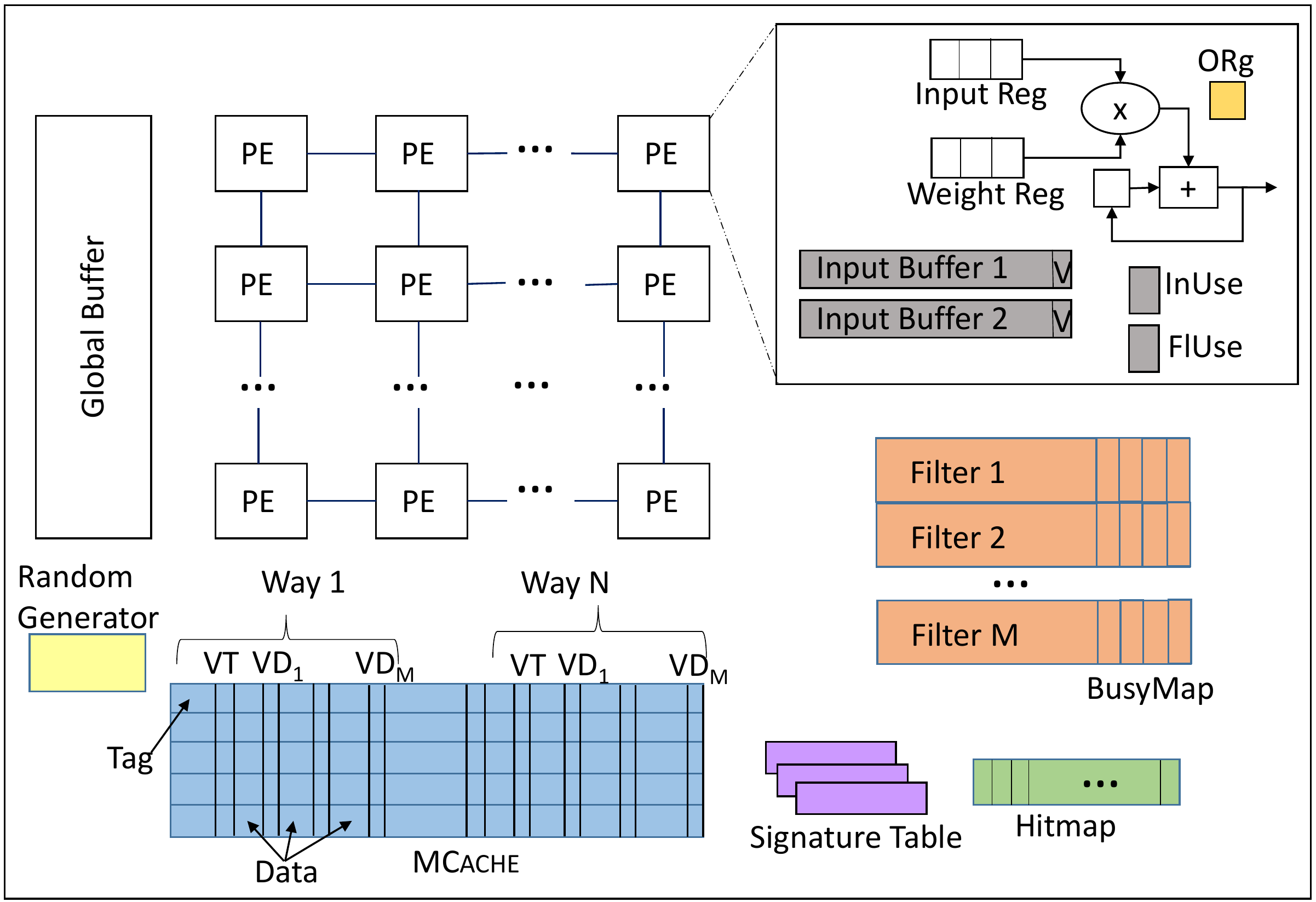}
\caption{Detailed modified design for the asynchronous design. Added structures are shown in color. PE-level changes allow a PE to operate on different input vectors, multiple filters allow different PE sets to operate on different filters, and the multi-version \scache\ allows to keep multiple filters' computations.}
\label{fig-modifications}
\vspace{-0.3cm}
\end{figure}
Figure~\ref{fig-modifications} shows the changes required for the asynchronous design. There are $3$ major changes. {\em First,} to store new input vectors each PE is extended to have two input buffers. Each buffer has an associated valid (V) bit to indicate whether it contains valid inputs or not. Moreover, each PE has a register, named {\bf InUse}, to indicate which of the two buffers is currently used. All PEs in a PE set will have the same value in the \textbf{InUse} register. With the extra buffer, whenever the fastest PE set completes computation, it loads next input vectors in streaming fashion as before. PEs in other PE sets store the new input vectors in the unused input buffer. That way, when those PE sets finish computations, new input vectors are already available in one of the input buffers. {\em Second,} the accelerator stores multiple filters in a shared buffer so that each PE can access it. Each filter has an associated \textbf{BusyMap} to indicate which PE sets are currently busy with that filter. Each PE maintains a register, named \textbf{FlUse}, to indicate which filter it is using. Like \textbf{InUse}, the \textbf{FlUse} register has the same value in each PE of a PE set. When all PE sets finish using a filter, it is loaded with a new filter and the \textbf{BusyMap} is initialized. {\em Third,} since each input vector and filter produces a new dot product result, we propose to make \scache\ a \textit{multi-version cache} where each cache line has multiple versions of data. Each data portion has a VD bit to indicate whether it is valid or not. Note that there are as many versions as the number of filters. Thus, if a PE set tries to use a new filter when there is no space to store it (because all M filters are marked busy with at least one PE set), then the PEs in the PE set will remain idle until a filter is completely used up in all PE sets. When the PEs in a PE set access a cacheline, they use \textbf{FlUse} register to determine which data version should be used.

\subsubsection{Input Similarity in Backward Propagation}
\label{sec-backward}
There are two major computations - (i) calculation of weight derivatives (${\bf dW_i}$) of the current layer (say, layer $i$), and (ii) calculation of output derivatives of the previous layer (${\bf dO_{i-1}}$). Here, boldfaced letters indicate multidimensional tensors. 
Now, let us consider the computation of ${\bf dO_{i-1}}$.  We observe that ${\bf O_i}$ is the same as ${\bf I_{i+1}}$. Therefore, if the filters of layer $i+1$ have the same dimension as those of layer $i$, the signatures and Hitmap produced by layer $i+1$ for $I_{i+1}$ can be applied to ${\bf dO_i}$ to find similarity. In that case, this computation scenario is the same as the forward propagation computation of layer $i+1$. Based on this observation, we propose saving signatures and the Hitmap of each layer during the forward propagation and reloading them during the previous layer's backward propagation. Then, \scheme\ applies the same technique as in $\S$~\ref{sec-forward}. However, if the filter's dimensions do not match, \scheme\ recalculates the signatures of gradient vectors and repopulates the Hitmap to reuse computations.

\begin{figure}[htpb]
\centering
\vspace{-0.4cm}
\includegraphics[width=0.7\columnwidth]{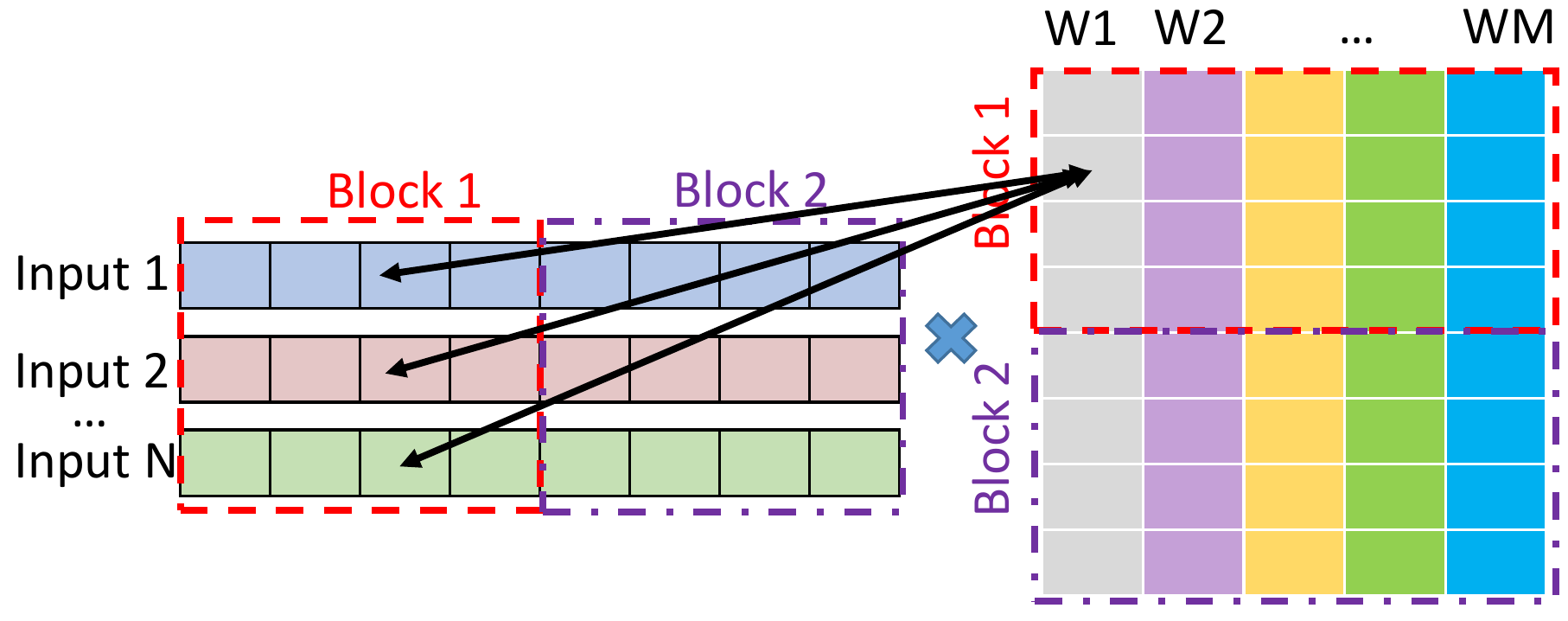}
\vspace{-0.2cm}
\caption{Computations in a fully connected layer. Input 1 to N forms a minibatch. The weight matrix has M columns.}
\label{fig-fullyconnected}
\vspace{-0.3cm}
\end{figure}

\subsubsection{Input Similarity in a Fully Connected Layer}
\label{sec-fc-clustering}
Figure~\ref{fig-fullyconnected} shows the high-level computations for a minibatch (size $N$) of inputs. 
Inputs and weights are divided into blocks based on the number of PEs. Suppose there are two input blocks and two weight blocks. 
One PE multiplies Input $1$ of block $1$ with $W_1$ of weight block 1 followed by $W_2$, $W_3$, ... $W_M$. Concurrently another PE multiplies Input $2$ and $W_1$ followed by $W_2$ to $W_M$. This continues up to Input $N$. Thus, if one input is similar to another, the earlier one's multiplication with $W_i$ ($1\le i\le M$) can be reused for the later one. Before the multiplication with weights begins, signatures and Hitmap are initialized based on the inputs of block 1 as in $\S$~\ref{sec-detect-local}. Before a PE multiplies an input vector (e.g., Input $2$) with a weight vector (e.g., $W_1$), the Hitmap of that input is checked. Depending on a HIT, MAU, or MNU, the result is either reused or calculated. This is similar to that of a convolution layer ($\S$~\ref{sec-forward}). 
In the case of a HIT, we store the id of the newly matched signature in the PE (that is processing the input corresponding to the stored signature causing the match). Let us refer to this PE as the earlier PE.
During the multiplication with weights, whenever the result for one weight gets ready, the earlier PE sends the result to later PEs (that are processing the inputs of the matched signatures) one by one to update their output. 
Sending results to other PEs is done in parallel with the regular operation. The earlier PE will start the operation of the next weight immediately after finishing with the current weight. In case the earlier PE is done with the next weight before the result is sent to all of the matched signature-related PEs, the earlier PE is stalled until the send operations finish. 
The earlier PE (after finishing block $1$ input) loads an input from block $2$ and starts signature generation while other PEs keep processing inputs and weights from block $1$. For doing this, we break the \scache\ into two separate caches, and PEs in each block will write and update the associated cache. The earlier PE can start multiplication with weights immediately after the signature generation process since the required results from block $1$ have already been computed and sent to the PEs that need it.
Race can arise when one PE wants to write to its related output memory while the earlier PE's results are ready, and it needs to write them into memory output too. 
We designed a simple conflict handler for handling these simultaneous write requests. 
Once all the inputs from block 1 is multiplied by the weights from block 1, PEs can start the operation of block 2 of inputs. Similar to block 1, they will start with signature generation and Hitmap initialization. Some PEs may have already started the operation of block 2 inputs, and their signatures are in the related \scache\ part, and new PEs can get results from these PEs in case of a signature match. 
\subsubsection{Input Similarity in an Attention Layer}
\label{sec-attention}
An attention layer is used in a sequence-to-sequence model~\cite{transformer}. 
Let's assume that the input vectors are $\mathbf{X^{t*k}={x_1, x_2, ..., x_t}}$ and the output vectors are $\mathbf{Y^{t*k}}={\mathbf{y_1}, \mathbf{y_2}, \ldots, \mathbf{y_t}}$. For simplicity, assume both $\mathbf{X}$ and $\mathbf{Y}$ to have the same sequence length (i.e., $t$) and the same vector representation of length $k$. To produce output vector $\mathbf{y_i}$, the attention layer simply takes a weighted average over all input vectors, $\mathbf{Y_j} = \sum_{j}W_{ij}X_j$. Here $\mathbf{W^{t*t}}$ is not parametric but rather a weighted matrix representing the correlation between each element of $\mathbf{X}$. $\mathbf{W}$ can be calculated as $\mathbf{W}=\mathbf{X*X^T}$. 
We can calculate $\mathbf{Y^{t*k}=W^{t*t}*X^{t*k}}$. Here $\mathbf{Y}$ is simply a matrix-matrix multiplication.
Thus, we can apply our idea of input similarity to calculate $\mathbf{Y}$ by exploiting the similarity among $\mathbf{x_i}$ vectors. Because this computation is similar to a fully connected layer, we apply the same technique as in $\S$~\ref{sec-fc-clustering}.

\vspace{-0.15cm}
\subsection{Adaptation in {\bf \scheme}}
\label{sec-adaptation}
\vspace{-0.05cm}
As training proceeds, DNN models become more sensitive to computation reuse. 
So, we propose \scheme\ to be adaptive. 

{\bf Increase in Signature Length: }
If signature length increases, vectors ${\bf v_1}$ and ${\bf v_2}$ are found to have similarity only when ${\bf v_1-v_2=\epsilon}$ becomes significantly smaller. So, a larger signature has a lesser effect on model accuracy but it may reduce computation reuse. Therefore, \scheme\ starts with a smaller signature size (e.g., 20 bits long) and progressively increases signature length as the model is trained more. Towards this end, \scheme\ calculates the average loss in each iteration during training. If there is no change in the loss for $K$ consecutive iterations, then \scheme\ increments signature length by 1. 

{\bf Stoppage of Similarity Detection: }
\scheme\ analytically determines if detecting similarity can save computations or not. If no computation can be saved, then \scheme\ turns off the similarity detection phase. In order to implement this, \scheme\ records the total computation cost (i.e., cycles) $C_S$ for signature generation in forward and backward propagation when some computations are reused. This cost is compared with the total computation cost $C_B$ of the baseline system without any computation reuse. $C_B$ can be calculated analytically using different hardware components' latency. If the former cost is more than the later for $T$ consecutive batches of inputs, \scheme\ stops generating signatures.

%% file: support.tex
\scheme\ can be easily implemented in other dataflow accelerators. Here we explain it for weight and input stationary dataflows.


{\bf Weight-Stationary Dataflow:} Weights are stationary in the PEs, and input vectors are broadcasted to PEs. 
\scheme\ keeps the original dataflow structure the same and loads random vectors as the first part of filters. So, random vectors will initially be loaded to different PEs, and one input vector will be broadcasted to PEs. The signatures are generated using the original Weight-Stationary dataflow.
However, after loading random vectors, one input vector's signature will be stored in several PEs, and several PEs will update the signature table for different vectors. Then, using the generated signature table and the proposed \scache\ structure, \scheme\ will detect similarity using signatures and specify the hit/miss for them. The next step is loading the regular filters, but \scheme\ already knows which input vectors are similar. So, while reading vectors from global memory, \scheme\ skips similar vectors and reuses the results.

{\bf Input-Stationary Dataflow:}
Inputs are stationary in PEs, and weights will be broadcasted. The hardware first finishes the operation of one input vector before loading a new input vector. The global structure of \scheme\ in Input-Stationary dataflow is similar to Weight-Stationary dataflow, which loads random vectors as the first part of filters. So, one input vector will be loaded to different PEs, random vectors will be loaded to different PEs at the beginning, and weight matrices will be streamed into PEs. 
The signatures are generated using the original Input-Stationary dataflow and the Hitmap is initialized accordingly.
During actual operation, if there is a hit for an input vector, \scache\ skips the rest of the weights and loads the next input vector. However, \scheme\ will continue streaming the weights if there is a miss for current input.

%% file: implement.tex

\label{sec-impl}

{\bf \scache\ Design:} To make \scache\ scalable, we make two design decisions.
{\em First}, \scache\ is implemented in shared memory using slice registers (flip flops in FPGA) that can be accessed through id. Multiple PE sets can read the same cache entry within a fixed delay using the id. That is why, 
the entry id is saved along with the signature in the signature table. Since an inserted signature is not removed from the cache until a new channel starts, further accesses to that signature are done through the id without requiring comparison.
{\em Second}, we add a queue and a simple controller for each cache set.
Thus, a cache set can be updated independently.
To insert multiple signatures simultaneously into a cache set,
the queue records the requests, and the controller serializes them - one at a time. Multiple signature insertions for different cache sets can proceed simultaneously without any issue. Although these techniques are specifically for FPGA, for an ASIC accelerator, similar techniques such as banked cache~\cite{hennessy}, multi-signature cache line, and PE set wise smaller cache can be used. 
We leave the details of ASIC design for future.

{\bf PE Implementation:} The structure of PE in \scheme\ is similar to a typical PE~\cite{eyeriss} as shown in Figure~\ref{fig-modifications}. Each PE has an extra block memory to store one input row. The original Row Stationary dataflow needs enough input buffers to store one row of one vector in each PE. This is implemented using Slice Registers which are limited in FPGA. In PE implementation, we reduce the input buffer to only one register and use the local block memory to process the inputs and store the results back.
For Synchronous design, since all PEs are in the same phase at a time, one extra block memory is enough. 
However, since the first and second phases can be done by different PE sets independently in Asynchronous design, we implement multiple filters using block memory. Table~\ref{tbl-types-memory} shows the types of memories used in the implementation of \scheme\ .
\begin{table}[htpb]
\centering
\vspace{-0.12cm}
\caption{Detailed memory types in \scheme\ design.\vspace{-0.2cm}}
\scalebox{0.7}{
\begin{tabular}{||l|l||} 
 \hline\hline
 Memory Type & \scheme\ Components \\
 \hline\hline
 Block Memory   & Global Buffer, Input Buffer, Signature Table \\
 \hline
 Slice Register & \scache, Filters, Hitmap, Input/Weight registers,\\
                & InUse/FIUse flags, ORg \\
 \hline\hline
\end{tabular}
}
\vspace{-0.2cm}
\label{tbl-types-memory}
\end{table}

%% file: experiment.tex
The hardware implementation of \scheme\ is done on a Virtex 7 FPGA board~\cite{virtex} configured using the Xilinx Vivado~\cite{vivado} software.
We used an Eyeriss-style~\cite{eyeriss} row stationary accelerator with the same number of PEs (i.e., 168) as the baseline. The model's inputs and weights are stored in an external SSD connected to the FPGA. We use 1024 entries and associativity of 16 for \scache. $\S$~\ref{sec-cache-size} analyzes the performance impact of different \scache\ organizations.
All performance, power consumption, hardware utilization, and other hardware-related metrics are collected from the synthesized FPGA design using Vivado. Different model accuracy results are collected from PyTorch~\cite{pytorch} implementation. 
We consider 12 networks: AlexNet, GoogleNet, VGG13, VGG16, VGG19, ResNet50, ResNet101, ResNet152, Inception-V4, MobileNet-V2, SqueezeNet-1-0, and Transformer. We use 80 image classes from ImageNet and report top 1\% accuracy for CNN models. For transformer, we use Multi30k dataset and report accuracy and Bleu score.
We train the models until the highest reported accuracy is reached in the baseline system. For \scheme, we trained the models with the same number of epochs. 
$\S$~\ref{experimental-scenarios} provides a comparison with UCNN~\cite{ucnn}, Unlimited Zero-Pruning, and Unlimited Similarity Detection.
\vspace{-0.2cm}

%% file: result.tex
\subsection{Accuracy and Performance Comparision}
\label{ovr-analysis}
\begin{figure}[htp]
\centering
\vspace{-0.35cm}
\includegraphics[width=0.65\columnwidth]{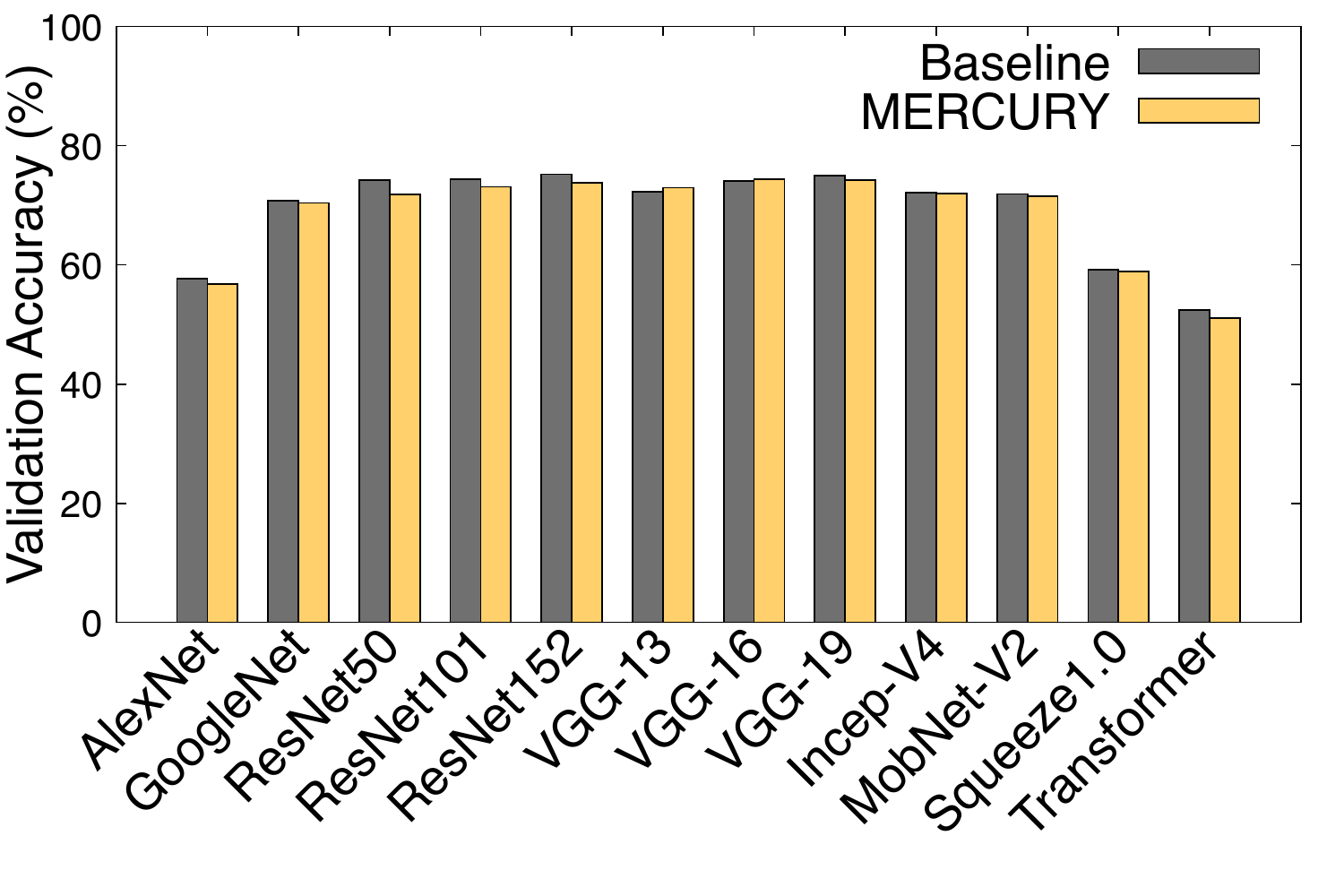}
\vspace{-0.4cm}
\caption{Validation accuracy for \scheme\ vs baseline.}
\label{acc-validate}
\vspace{-0.4cm}
\end{figure}

\vspace{-0.05cm}
Figure~\ref{acc-validate} shows the impact of \scheme\ on the models' accuracy. Overall, there is a $0.7\%$ reduction in validation accuracy\footnote{Hyper-parameters are adjusted in training to achieve optimal accuracy}. Considering the inherent randomness during the training process, we argue that \scheme's accuracy is comparable to the baseline system. Also, we got the same bleu\_score of $33.52$ on test data for the transformer. 

\begin{figure*}[!htpb]
\centering
\begin{subfigure}{0.32\textwidth}
    \centering
    \includegraphics[width=\columnwidth]{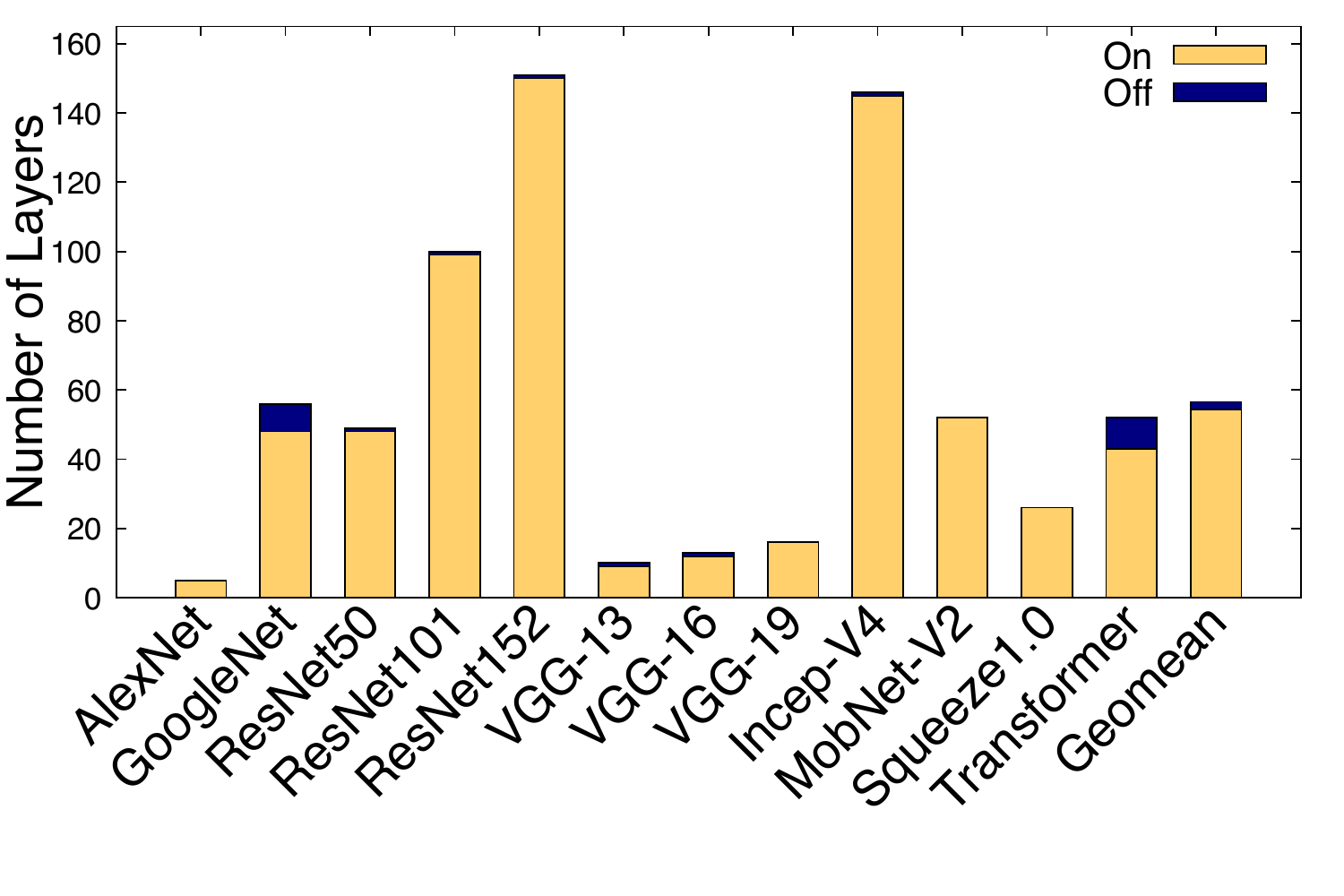}
    \vspace{-0.8cm}
    \caption{Adaptivity across layers of models}
    \label{detection-mode}
\end{subfigure}
\begin{subfigure}{0.32\textwidth}
    \centering
    \includegraphics[width=\columnwidth]{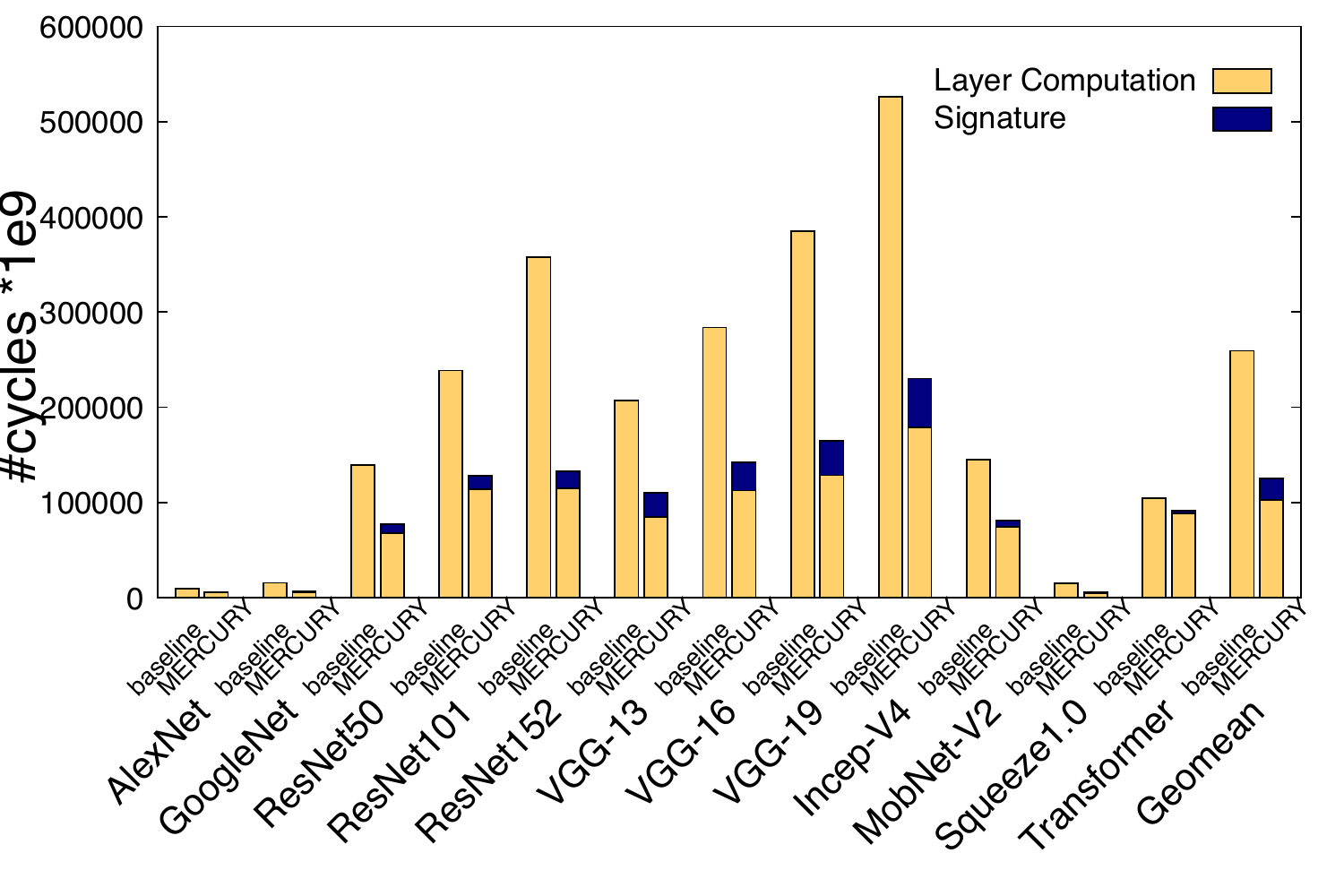}
    \vspace{-0.8cm}
    \caption{Computational cycle breakdown}
    \label{ovr-savings}
\end{subfigure}
\begin{subfigure}{0.32\textwidth}
    \centering
    \includegraphics[width=\columnwidth]{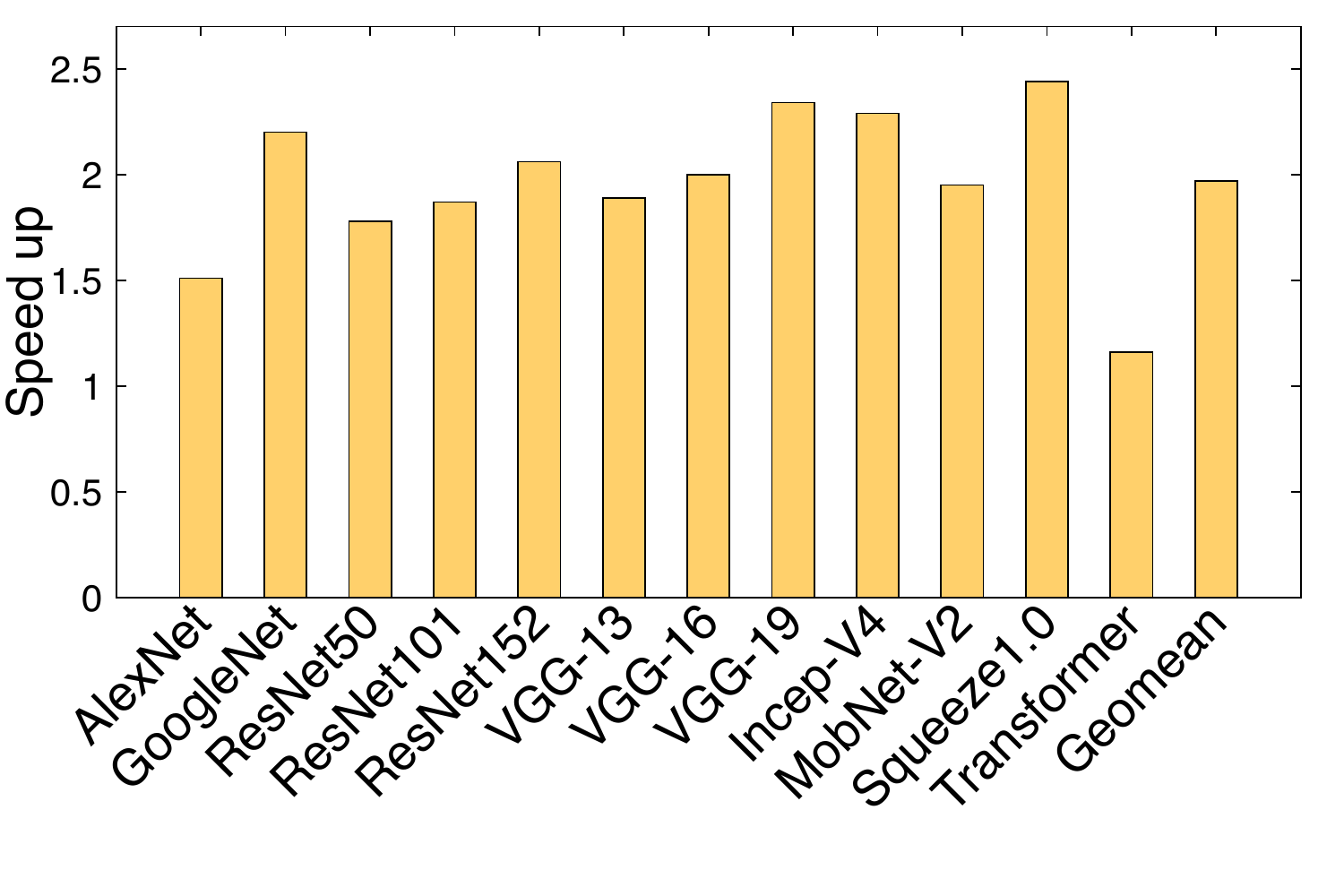}
    \vspace{-0.8cm}
    \caption{Speed up}
    \label{speedup}
\end{subfigure}
\vspace{-0.2cm}
\caption{\scheme\ performance with respect to the baseline.}
\vspace{-0.4cm}
\label{fig-ovr-perf}
\end{figure*}

Figure~\ref{detection-mode} shows the adaptivity of \scheme\ across different models. Based on the similarity detection costs in layers and the overall performance during the model training, \scheme\ may turn off similarity detection in some layers during the training. Figure~\ref{ovr-savings} shows the computational cycle breakdown of \scheme\ and baseline. This cycles includes computation cycles of convolutional, fully connected, and attention layers as well as signature calculation. Most of the cycles belong to the layer computations. The signature computation accounts for only a fraction of the total cycles. Overall, \scheme\ can reduce the total computation time by about $50\%$, which results in an average speedup of $1.97\times$, as shown in Figure~\ref{speedup}. For bigger networks, such as ResNet152, VGG19, and Inception-V4, there are more saving opportunities since there are more chances of similarity between vectors. 

\vspace{-0.15cm}
\subsection{Case Study: VGG13}
\label{detail-analysis}
\begin{figure*}[!htpb]
\centering
    \begin{subfigure}{0.32\textwidth}
    \centering
    \includegraphics[width=\columnwidth]{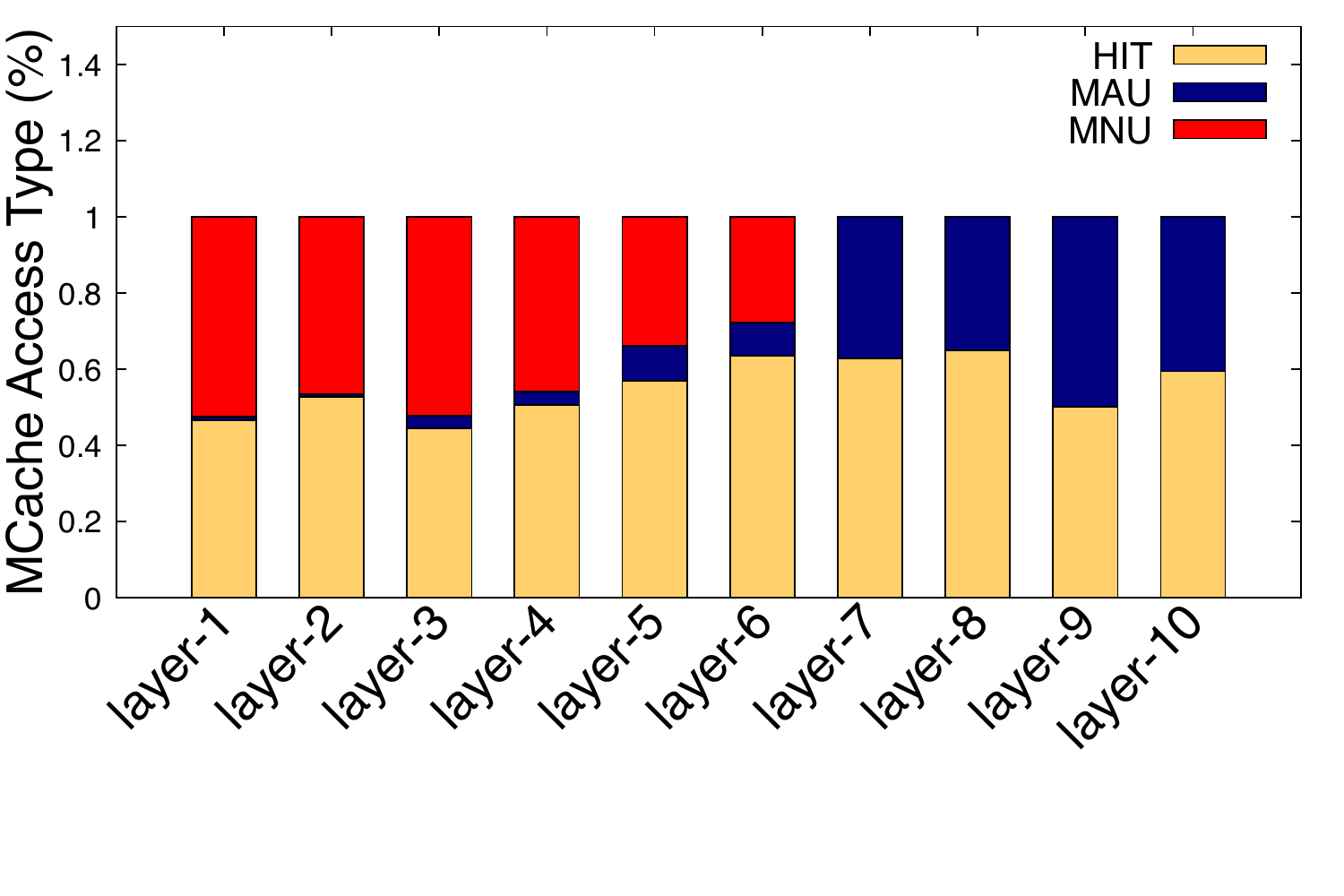}
    \vspace{-0.8cm}
    \caption{Characterization of \scache\ access}
    \label{cache-scache-access}
    \end{subfigure}
    \begin{subfigure}{0.32\textwidth}
    \centering
    \includegraphics[width=\columnwidth]{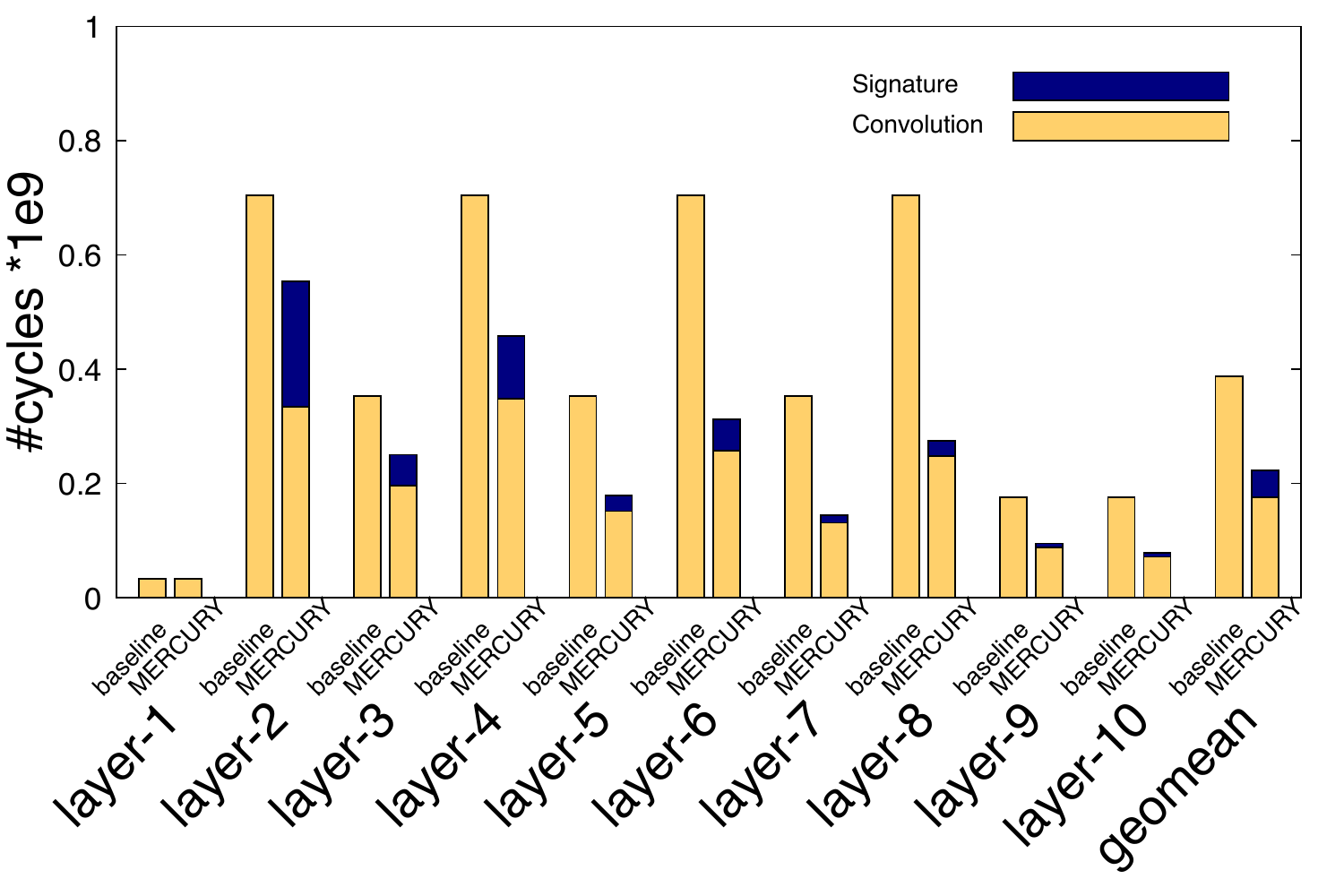}
    \vspace{-0.8cm}
    \caption{\footnotesize{Cycles across layers}}
    \label{cycle-save-layer} 
    \end{subfigure}
    \begin{subfigure}{0.32\textwidth}
    \centering
    \includegraphics[width=\columnwidth]{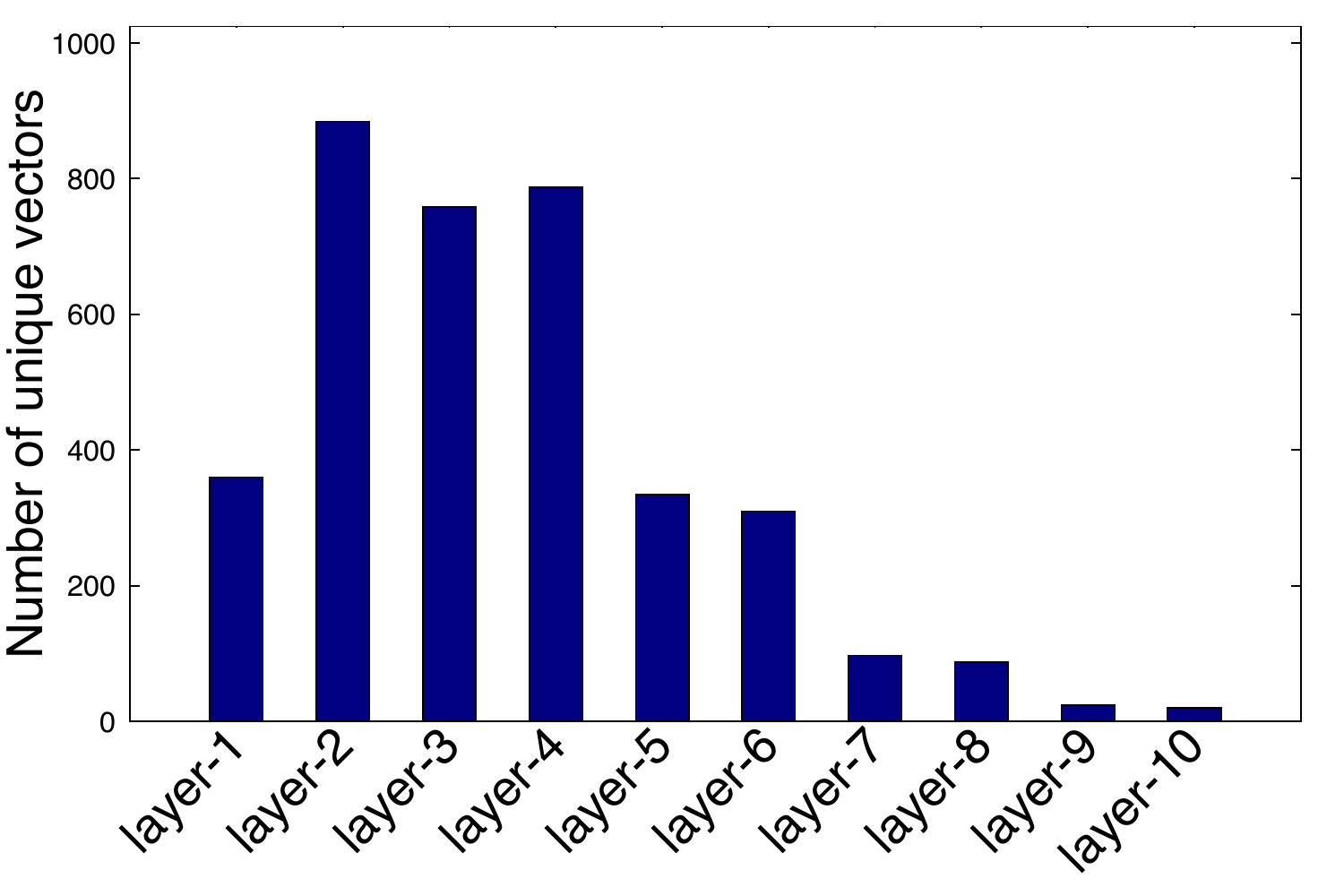}
    \vspace{-0.8cm}
    \caption{\footnotesize{Number of unique vectors across layers}}
    \label{clusters-per-layer}
    \end{subfigure}
    \vspace{-0.2cm}
    \caption{Characterization of \scheme\ in VGG13.}
    \label{vgg13-detail}
    \vspace{-0.65cm}
\end{figure*}

We conduct a detailed analysis of VGG13 to show how \scheme\ works at runtime, focusing on the characterization of the \scache\ access, the savings, and the number of unique vectors found across layers. The results are shown in Figure~\ref{vgg13-detail}. Figure~\ref{cache-scache-access} shows a gradual increase in \scache\ Hit and MAU percentage due to the reduction in the number of input vectors and cache occupants.
Figure~\ref{cycle-save-layer} shows that the computational cycles may vary across layers of VGG13 due to the difference in layers' size and channels. Some layers have a higher cycle count related to the input size and number of input and output channels, and the amount of savings differs. The number of unique vectors is also different across layers. As shown in Figure~\ref{clusters-per-layer}, the first few layers have the highest number of unique vectors since they have a large input size. This value is lower for the later layers due to the smaller input size.

\subsection{Performance Impact of Different Organizations of \scache}
\label{sec-cache-size}
\begin{figure*}[!htpb]
\centering
    \begin{subfigure}{0.32\textwidth}
    \centering
    \includegraphics[width=\columnwidth]{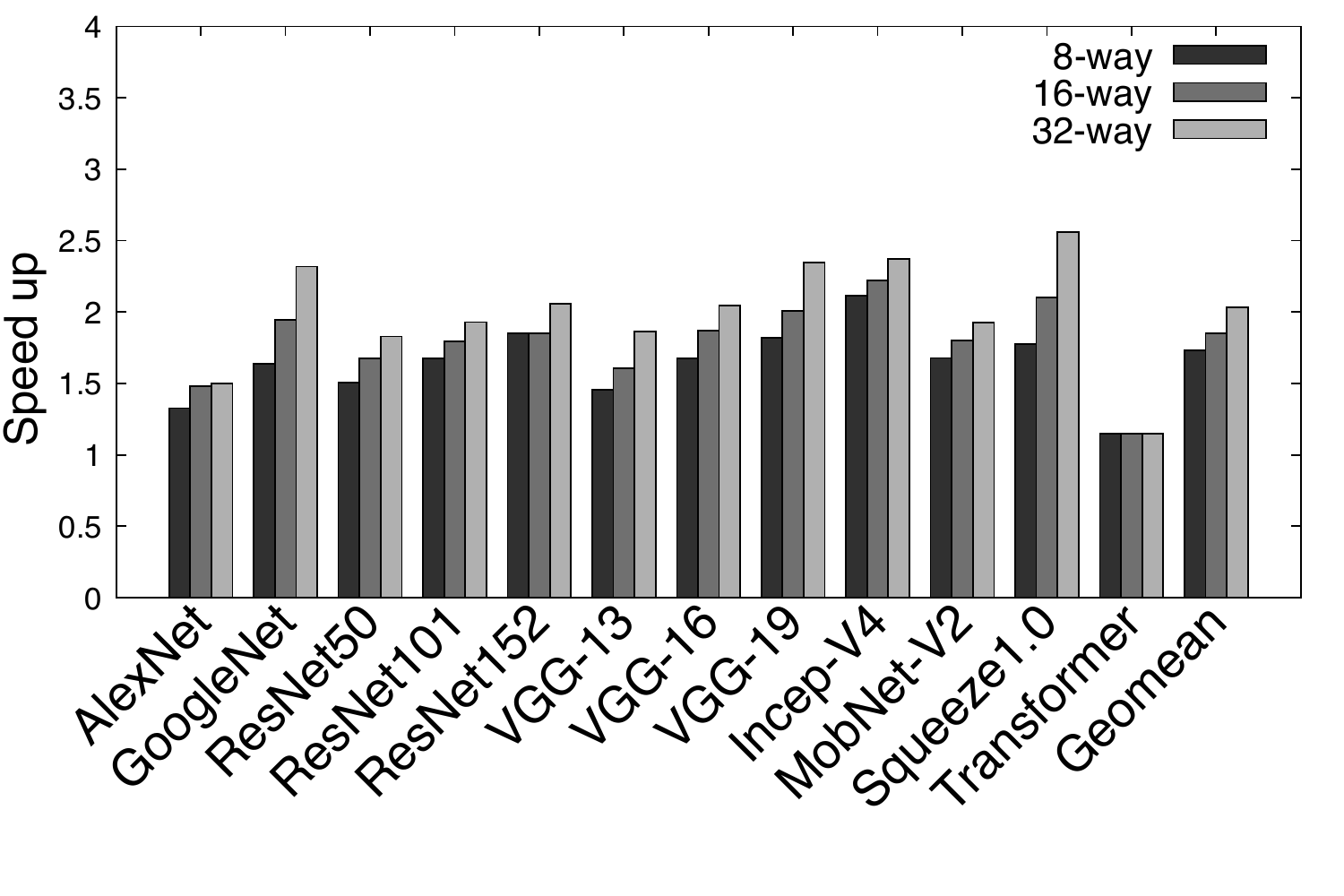}
    \vspace{-1cm}
    \caption{\footnotesize{Cache size = 512 entries.}}
    \end{subfigure}
    \begin{subfigure}{0.32\textwidth}
    \centering
    \includegraphics[width=\columnwidth]{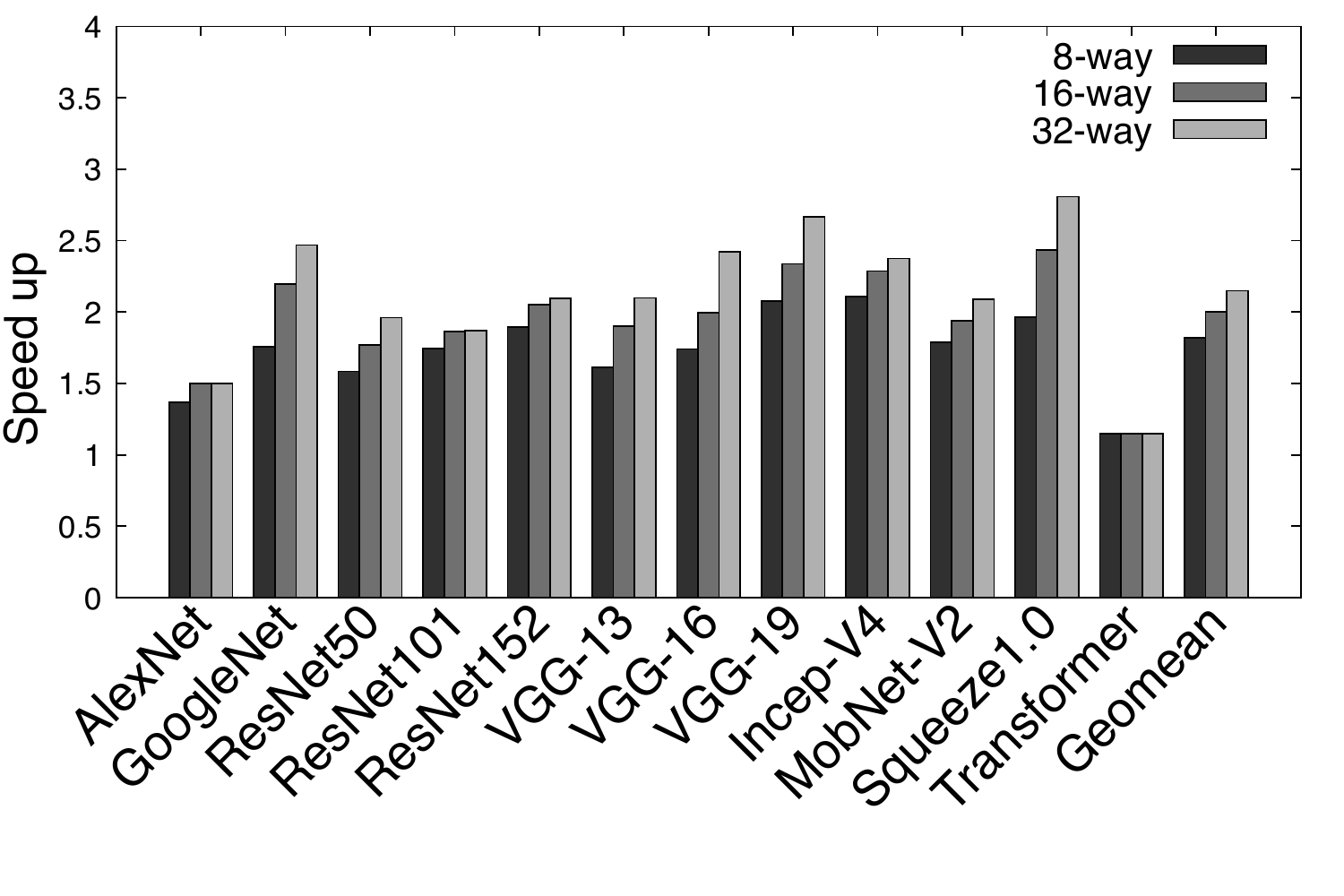}
    \vspace{-1cm}
    \caption{\footnotesize{Cache size = 1024 entries.}}
    \end{subfigure}
    \begin{subfigure}{0.32\textwidth}
    \centering
    \includegraphics[width=\columnwidth]{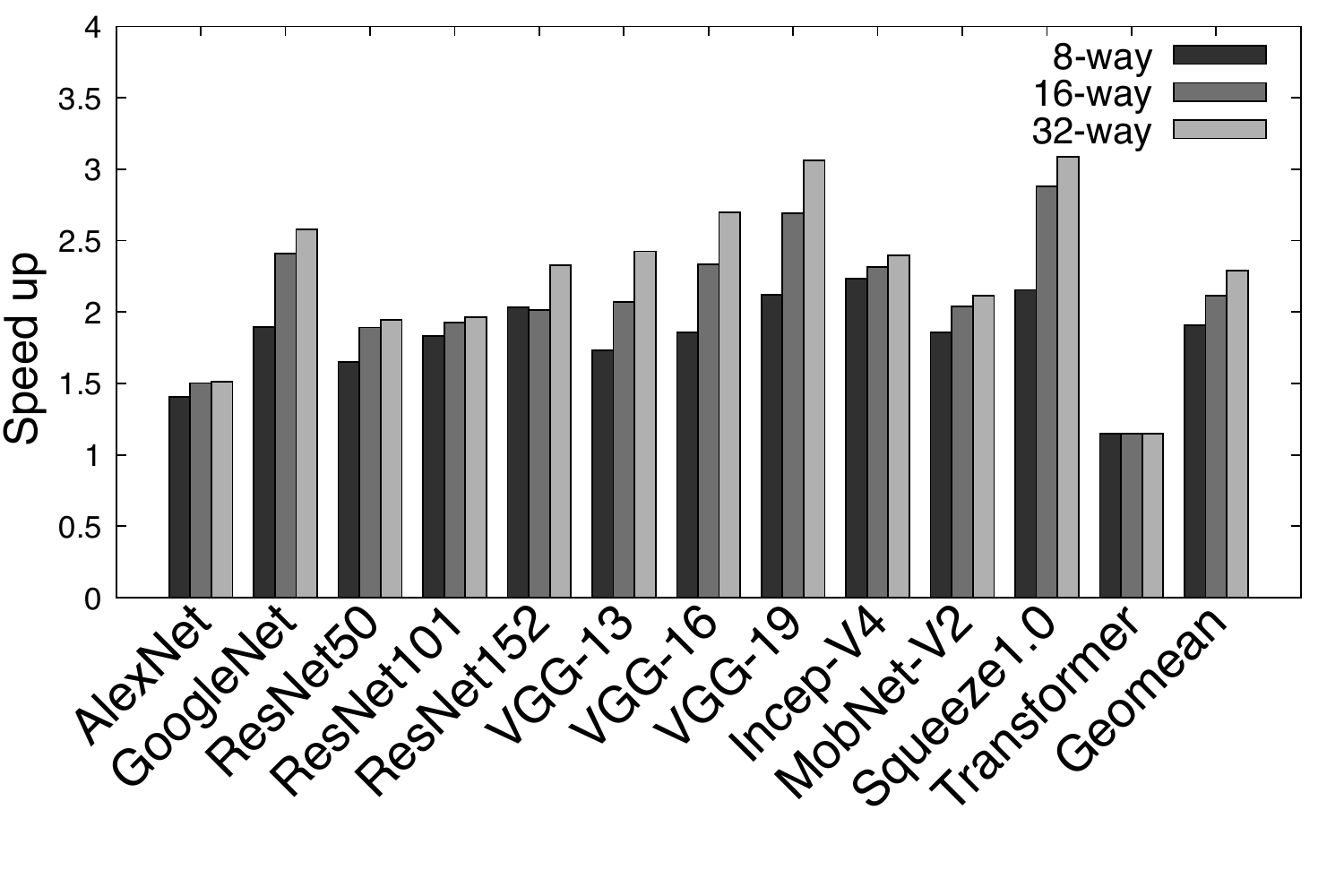}
    \vspace{-1cm}
    \caption{\footnotesize{Cache size = 2048 entries}}
    \end{subfigure}
    \vspace{-0.2cm}
    \caption{Impact of \scache\ organizations on the performance of \scheme.}
    \vspace{-0.5cm}
    \label{cache-sz-comp}
\end{figure*}

This section analyzes the impact of different \scache\ sizes and organizations on the performance of \scheme. As shown in Figure~\ref{cache-sz-comp}, \scheme\ performance increases with cache size and associativity. Unfortunately, Vivado did not finish synthesis of 32-way cache-based designs even after one day (our time limit for the server).
Thus, we had to limit ourselves to 8 or 16-way. Combining with Table~\ref{same-set-diff-way} shows that moving from a 512-entry, 8-way to a 1024-entry, 16-way cache only increases the power consumption by $2.85\%$ while giving $4.88\%$ more speed up. Doubling the cache size to $2048$ entries gives insignificant performance gains; thus, $1024$-entry and $16$-way is selected as the default configuration for \scache.

\subsection{Comparative Analysis}
\label{experimental-scenarios}

\begin{figure*}[!htpb]
\centering
    \begin{subfigure}{0.32\textwidth}
    \centering
    \includegraphics[width=\columnwidth]{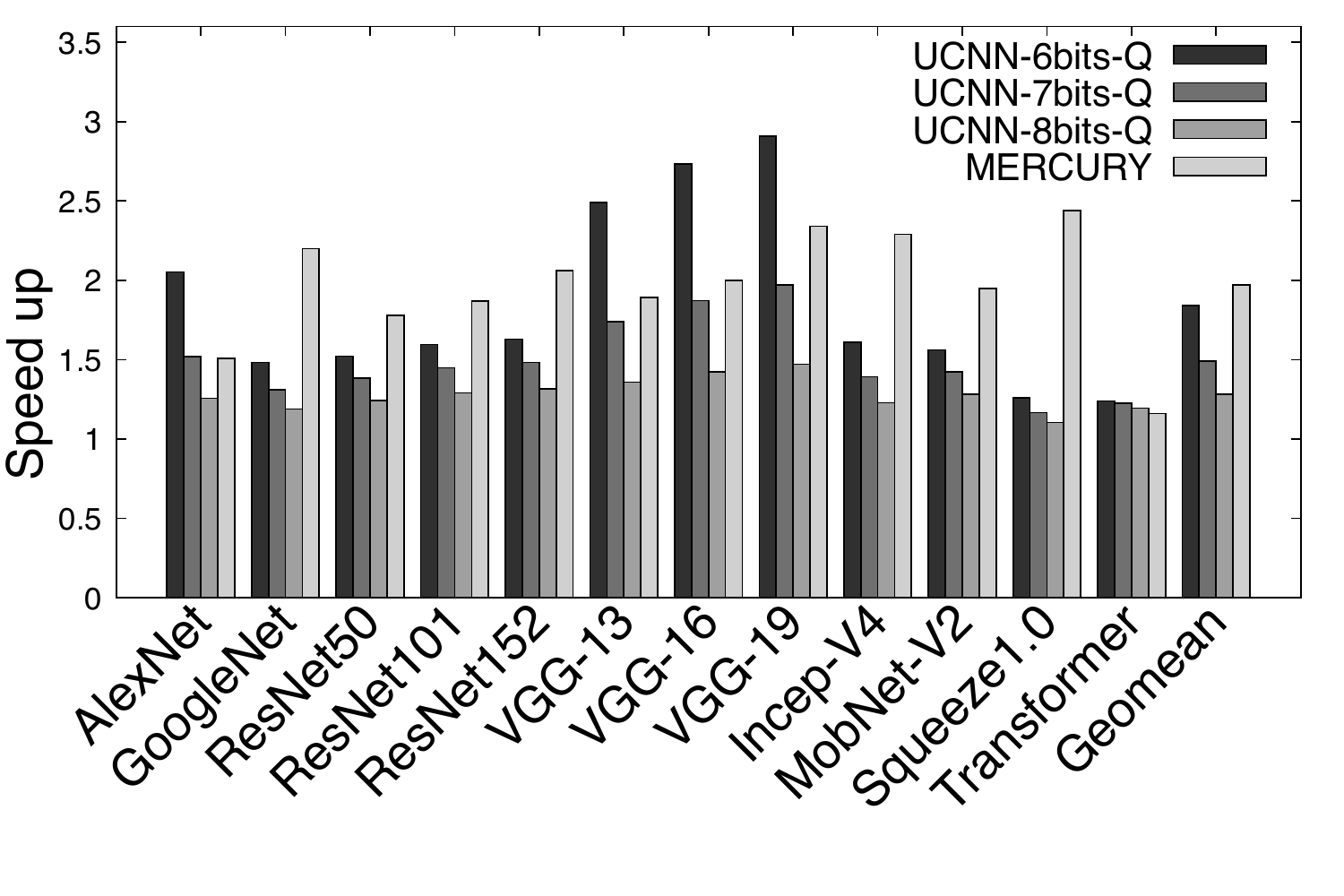}
    \vspace{-0.9cm}
    \caption{UCNN}
    \label{ucnn-speedup}
    \end{subfigure}
    \begin{subfigure}{0.32\textwidth}
    \centering
    \includegraphics[width=\columnwidth]{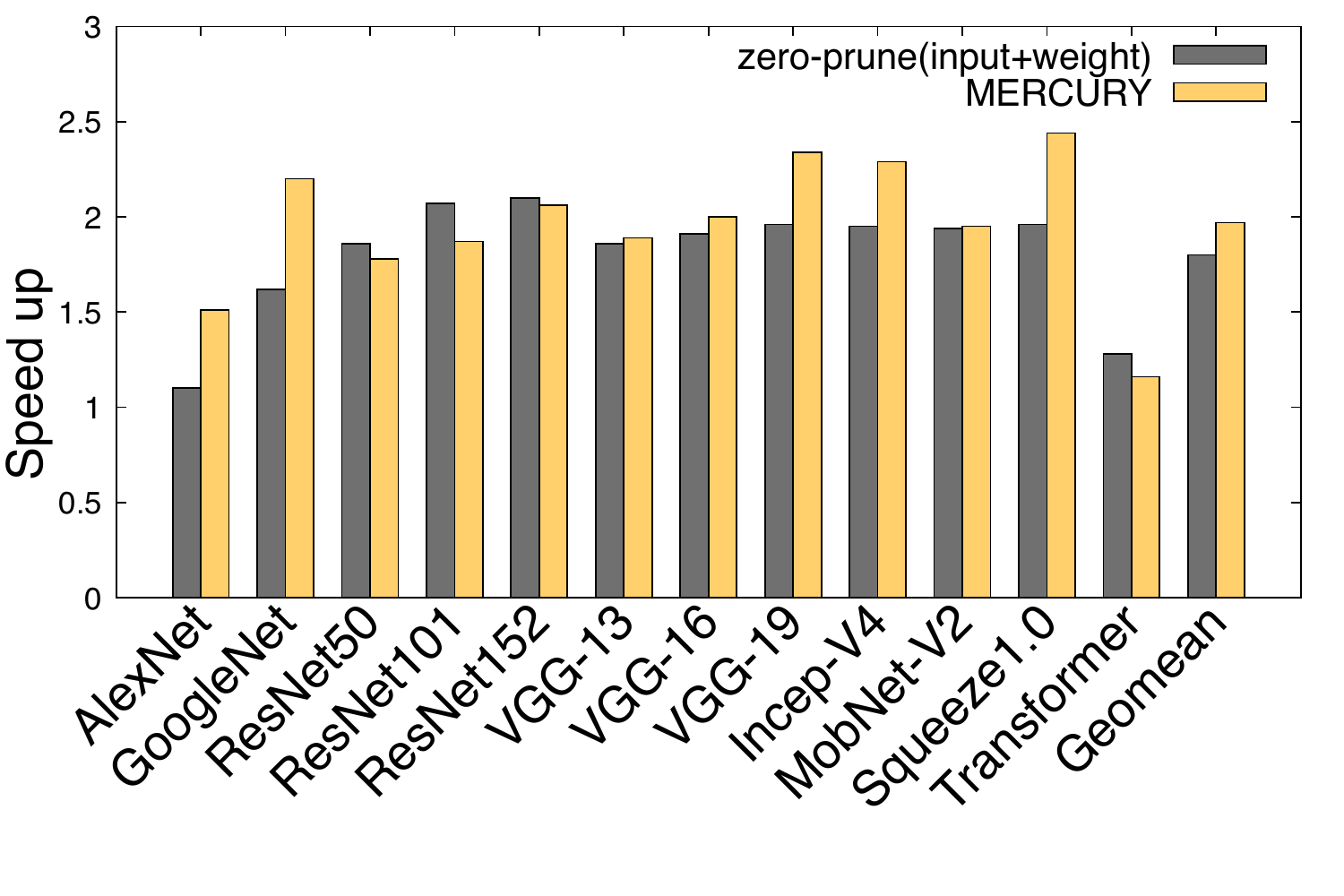}
    \vspace{-0.9cm}
    \caption{Unlimited zero pruning}
    \label{zprune-speedup}
    \end{subfigure}
    \begin{subfigure}{0.32\textwidth}
    \centering
    \includegraphics[width=\columnwidth]{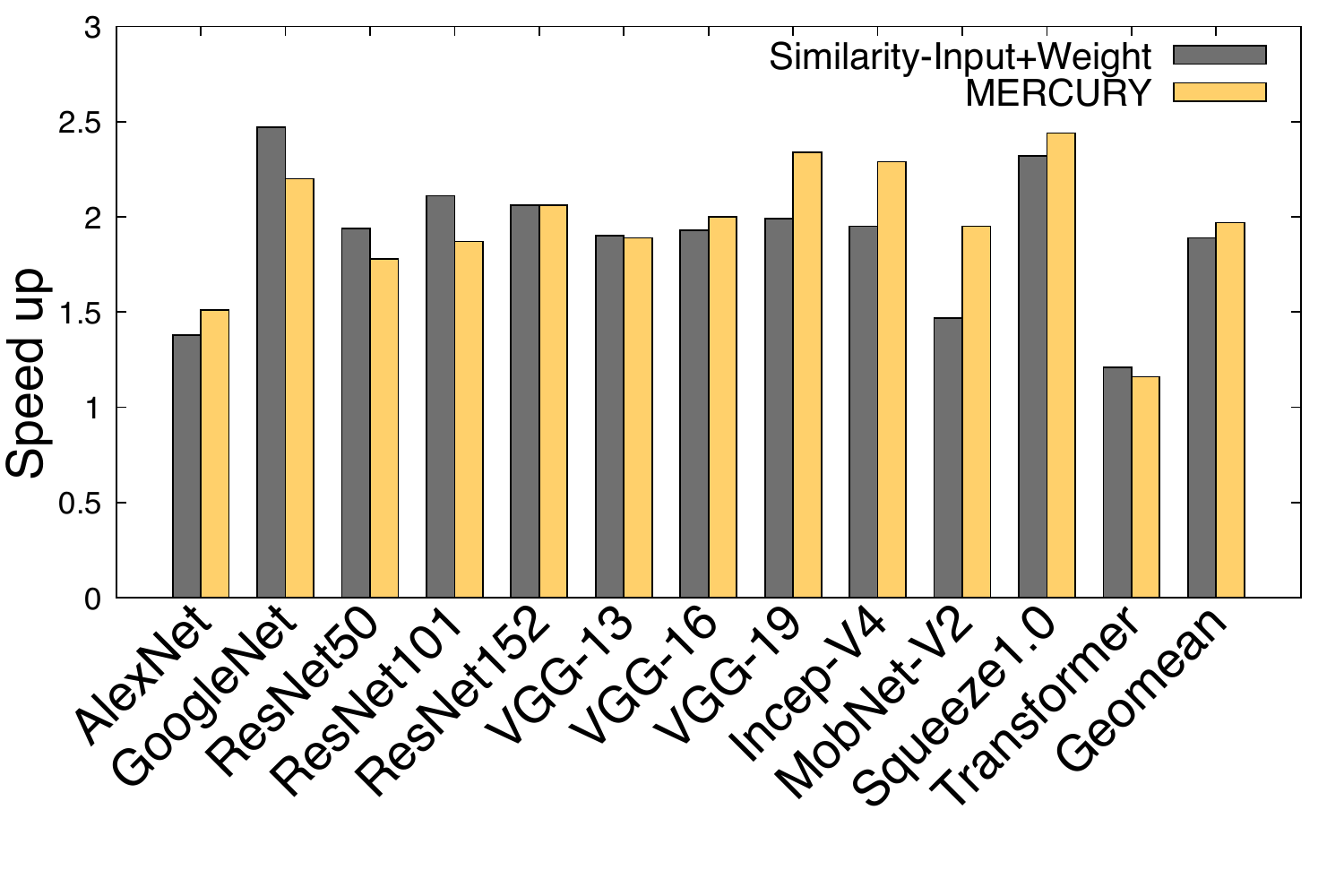}
    \vspace{-0.9cm}
    \caption{Unlimited similarity detection}
    \label{in-w-sim-speedup}
    \end{subfigure}
    \vspace{-0.2cm}
    \caption{Speed up of \scheme\ and other techniques.} 
    \vspace{-0.1cm}

    \label{fig:ucnn}
\end{figure*}

\subsubsection{Comparison with UCNN in inference mode}
\label{compare-state-of-the-art}
To compare with UCNN during inference, we perform the following scenario. {\em First}, we compute the static quantization for all layers of the pretrained model with a different number of bits. We only considered 6 to 8-bit quantizations since there was a significant loss in accuracy beyond this point. With 6-bits quantization, the final accuracy drops about $3\%$. {\em Second}, based on~\cite{ucnn}, we need to find similar items in one filter, do the activation group reuse, and then save all the additions with totally or partially similar activation groups. Due to the lack of access to the implementation of~\cite{ucnn}, 
we considered the maximum saving in each filter. In other words, in an activation group reuse with length \textit{k}, suppose we have similar indices in other filters, and we can save additions between all members. 
Figure~\ref{ucnn-speedup} shows a comparison between \scheme\ and the maximum achievable speedup of UCNN with different quantization policies. On average, \scheme\ outperforms UCNN with 7 and 8-bit quantization while achieving comparable performance gains to the 6-bit quantization version.

\subsubsection{Comparison with Unlimited Zero Pruning}
\label{compare-zero-pruning}
Here, we compare \scheme\ against the theoretical upper bound of Zero-Pruning, which assumes the accelerator can detect and save all zero-related computations in both input and weights. Figure~\ref{zprune-speedup} shows the comparison between \scheme\ versus the theoretical maximum achievable speedup of Zero Pruning. On average, \scheme\ outperforms by 4\%. Note that in actual hardware, the performance gains of Zero Pruning will be limited by hardware resource constraints and zero-value detection and bypass overhead.

\subsubsection{Comparison with Unlimited Similarity Detection}
\label{compare-in-w-sim}
This scenario finds and saves all similar elements in a model's inputs and weights. Similar to the Zero Pruning scenario, we did not consider any limitations on the amount of similarity. We assumed that the accelerator could find and save the computation of all similar elements. Figure~\ref{in-w-sim-speedup} shows the speedup of \scheme\ versus the Unlimited Similarity Detection scheme. On average, our approach performs $2\%$ better than the Unlimited Similarity technique. 

\subsection{Results with Other Dataflows}
\label{exp:other-dataflow}
\begin{figure}[h]
\centering
 \vspace{-0.1cm}
\begin{subfigure}{0.48\columnwidth}
\centering
\includegraphics[width=\columnwidth]{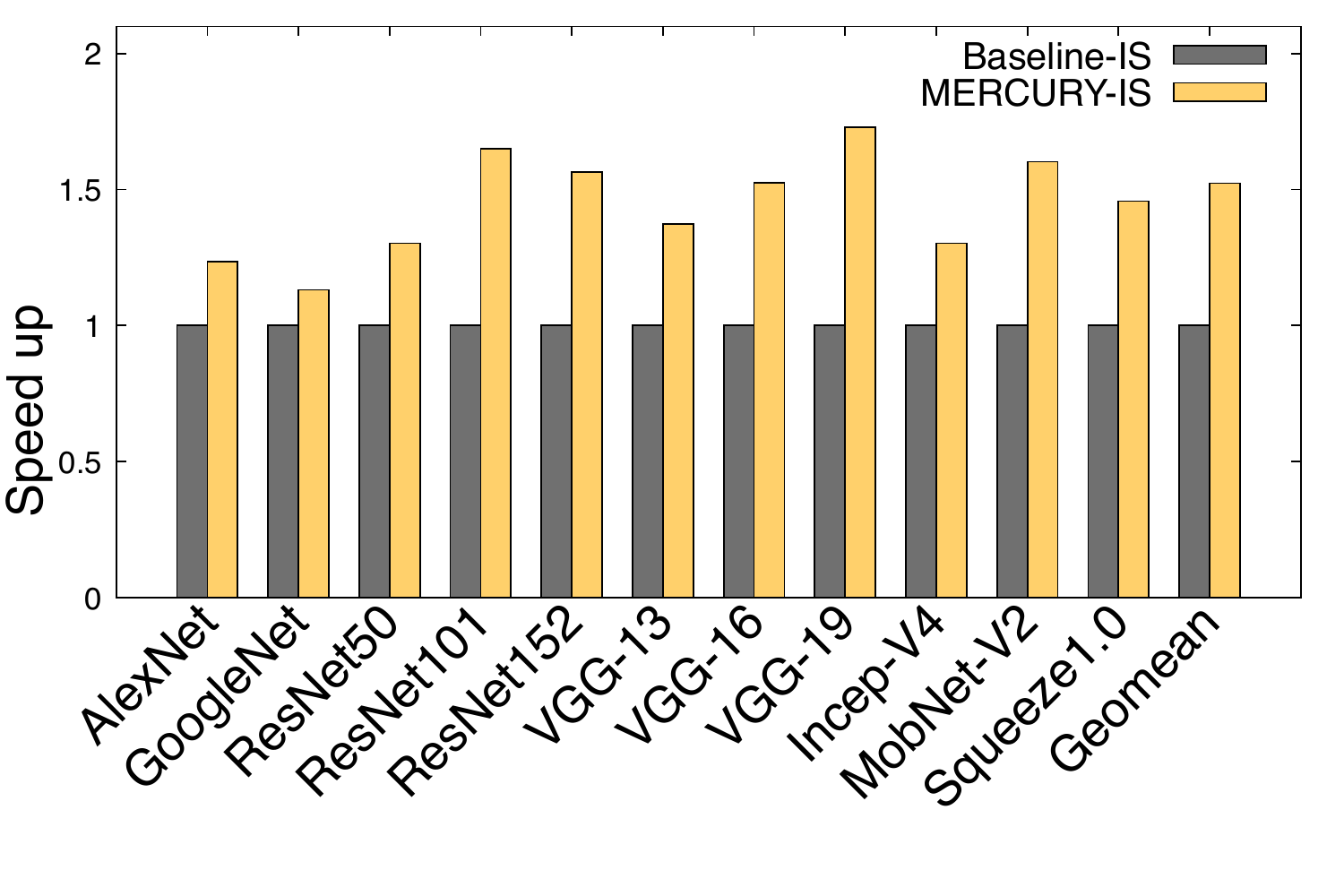}
\vspace{-0.85cm}
\caption{Input-stationary dataflow}
\label{IS-speedup}
\end{subfigure}
\begin{subfigure}{0.48\columnwidth}
\centering
\includegraphics[width=\columnwidth]{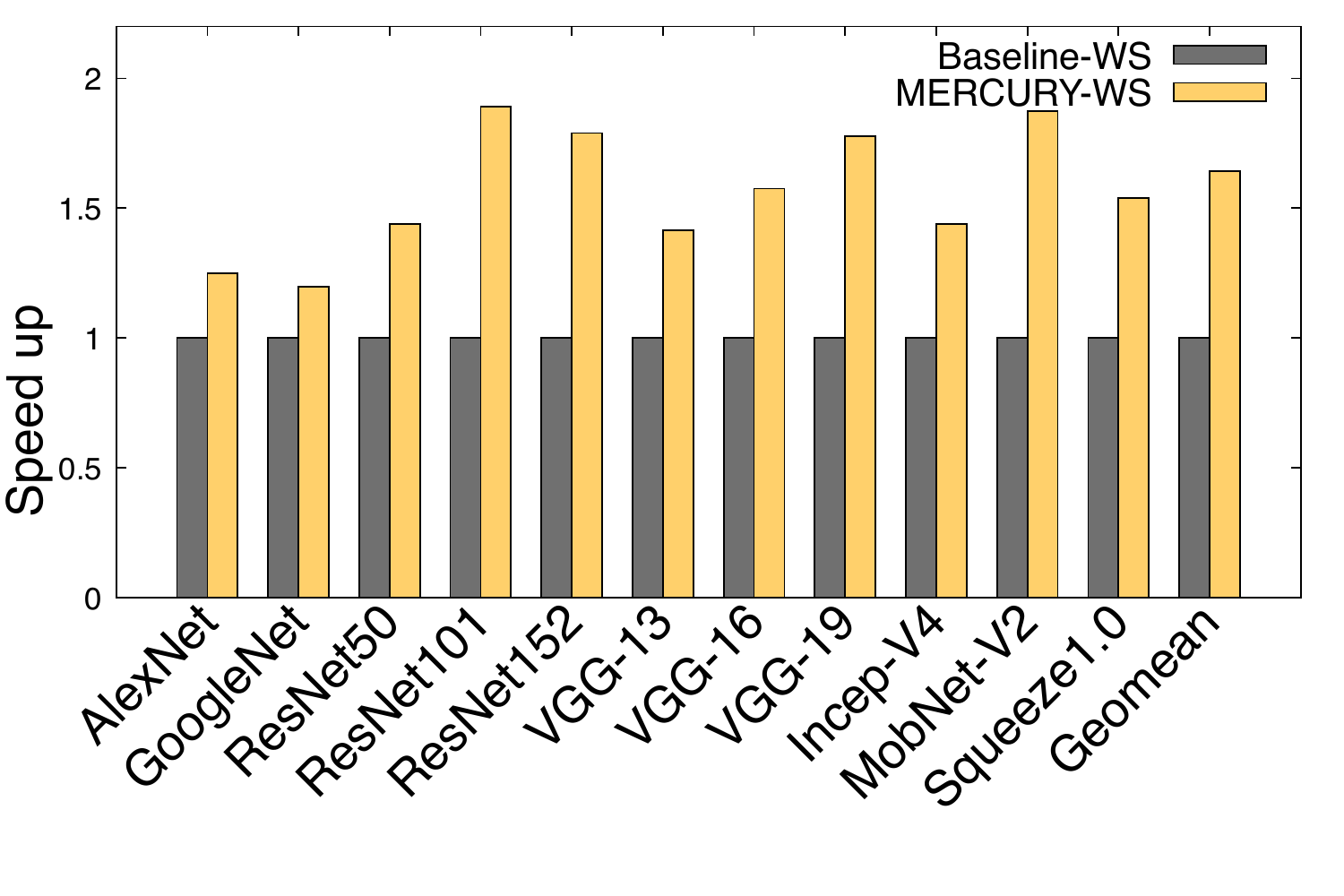}
\vspace{-0.85cm}
\caption{Weight-stationary dataflow}
\label{WS-speedup}
\end{subfigure}
\vspace{-0.15cm}
\caption{\scheme\ with different dataflows.}
 \vspace{-0.7cm}
\label{fig-dataflow}
\end{figure}

Figure~\ref{IS-speedup} and~\ref{WS-speedup} show the speedup of our scheme when deployed on top of the Input-Stationary and Weight-Stationary dataflow. For Input-Stationary, \scheme\ gives an average performance gains of $1.55\times$ over the baseline. The maximum speedup of $1.72\times$ is achieved in VGG-19. \scheme\ works better with Weight-Stationary as it achieves an average speedup of $1.66\times$ over the baseline. The maximum speedup of $1.89\times$ is observed in ResNet101.

\subsection{Hardware Utilization Analysis}
\label{hardware-analysis}

\begin{table*}
\footnotesize
    \begin{minipage}[t]{0.32\linewidth}\centering
    \scriptsize
    \caption{a) Resource usage and b) On-Chip Power Consumption (watt) Comparison of \scheme\ vs baseline for number of ways=16 and different set size.}
    \vspace{-0.1cm}
    \scalebox{0.7}{
    \begin{tabular}{||c|c||c|c|c|c||} 
     \hline\hline
     Cache & \# & \#Slice & \#Slice & \#Block &  \\
     Size & Sets & LUTs & Registers & RAM & \#DSP48E1s \\
     \hline\hline
    256 & 16 & 140597 & 62620 & 1177.5 & 198 \\\hline
    512 & 32 & 211437 & 69536 & 1193.5 & 198 \\\hline
    768 & 48 & 216544 & 74925 & 1209.5 & 198 \\\hline
    1024 & 64 & 216918 & 81332 & 1225.5 & 198 \\\hline
     \hline
    \end{tabular}}
    \begin{center} 
    (a) Resource usage
    \vspace{-0.1cm}
    \end{center}
    \scalebox{0.7}{
    \begin{tabular}{||c||c|c|c|c|c|c|c||} 
     \hline\hline
    \# &  &  &  & Block & & & \\
    Sets &  Clocks & Logic & Signals & RAM & DSPs & Static & Total \\
     \hline\hline
    16 & 0.138 & 0.102 & 0.18 & 0.516 & 0.087 & 0.681 & 1.811 \\\hline
    32 & 0.154 & 0.104 & 0.175 & 0.524 & 0.087 & 0.683 & 1.833 \\\hline
    48 & 0.155 & 0.103 & 0.201 & 0.548 & 0.087 & 0.685 & 1.884 \\\hline
    64 & 0.166 & 0.105 & 0.216 & 0.561 & 0.087 & 0.687 & 1.929 \\\hline
     \hline
    \end{tabular}}
    \begin{center} 
    (b) On-chip power consumption (watt)
    \end{center}
    \vspace{-0.1cm}
    \label{same-way-diff-set}
    \end{minipage}
    \hfill
    \begin{minipage}[t]{0.32\linewidth}\centering
    \scriptsize
    \caption{a) Resource usage and b) On-chip power consumption (watt) of \scheme\ vs baseline for set size=64 and different number of ways.}
    \vspace{-0.1cm}
    \scalebox{0.7}{
    \begin{tabular}{||c|c||c|c|c|c||} 
     \hline\hline
     Cache &  &\#Slice & \#Slice & \#Block &  \\
     Size & \#Ways & LUTs & Registers & RAM & \#DSP48E1s \\
     \hline\hline
    128 & 2 & 216777 & 65727 & 1225.5 & 198 \\\hline
    256 & 4 & 216618 & 67897 & 1225.5 & 198 \\\hline
    512 & 8 & 216758 & 71999 & 1225.5 & 198 \\\hline
    1024 & 16 & 216918 & 81332 & 1225.5 & 198 \\\hline
     \hline
    \end{tabular}}
    \begin{center} 
    (a) Resource usage
    \vspace{-0.1cm}
    \end{center}
    \scalebox{0.7}{
    \begin{tabular}{||c||c|c|c|c|c|c|c||} 
     \hline\hline
     \# &  &  &  & Block & & & \\
     Ways & Clocks & Logic & Signals & RAM & DSPs & Static & Total \\
     \hline\hline
    2 & 0.146 & 0.1 & 0.176 & 0.555 & 0.087 & 0.686 & 1.855 \\\hline
    4 & 0.151 & 0.104 & 0.197 & 0.543 & 0.087 & 0.686 & 1.874 \\\hline
    8 & 0.157 & 0.101 & 0.18 & 0.559 & 0.087 & 0.686 & 1.876 \\\hline
    16 & 0.166 & 0.105 & 0.216 & 0.561 & 0.087 & 0.687 & 1.929 \\\hline
     \hline
    \end{tabular}}
    \begin{center} 
    (b) On-chip power consumption (watt)
    \end{center}
    \vspace{-0.1cm}
    \label{same-set-diff-way}
    \end{minipage}
    \hfill
    \begin{minipage}[t]{0.32\linewidth}\centering
    \scriptsize
    \caption{a) Resource usage and b) On-Chip Power Consumption (watt) Comparison of \scheme\ vs baseline (cache size=1024 and number of ways=16).}
    \vspace{-0.1cm}
    \vspace{0.3cm}
    \scalebox{0.7}{
    \begin{tabular}{||c||c|c|c|c|c|c||} 
     \hline\hline
      & \#Slice & \#Slice & \#Block  & \\
     Method & LUTs & Registers & RAM & \#DSP48E1s \\
     \hline\hline
    Baseline & 56910 & 48735 & 1161.5 & 198 \\\hline
    \scheme\ & 216918 & 81332 & 1225.5 & 198 \\\hline
     \hline
    \end{tabular}}
    \begin{center} 
    (a) Resource usage
    \vspace{0.3cm}
    \end{center}
    \scalebox{0.6}{
    \begin{tabular}{||c||c|c|c|c|c|c|c||} 
     \hline\hline
      &  &  &  & Block & & & \\
     Method & Clocks & Logic & Signals & RAM & DSPs & Static & Total \\
     \hline\hline
    Baseline & 0.112 & 0.07 & 0.138 & 0.511 & 0.087 & 0.678 & 1.703 \\\hline
    \scheme\ & 0.166 & 0.105 & 0.216 & 0.561 & 0.087 & 0.687 & 1.929 \\\hline
     \hline
    \end{tabular}}
    \begin{center} 
    (b) On-chip power consumption (watt)
    \end{center}
    \vspace{-0.1cm}
    \label{skipcnn-vs-baseline}
    \end{minipage}
\vspace{-0.6cm}
\end{table*}
In this section, we provide a detailed analysis of the resource usage and power consumption of \scheme\ implementation on the Virtex 7 FPGA board.
Table~\ref{same-way-diff-set} compares the resource usage and on-chip power consumption of \scheme\ for different \scache\ sizes. Specifically,  quadrupling the number of \scache\ sets only increases the overall power consumption by $6.5\%$. If we fix the number of sets to $64$ and increase the number of ways from $2$ to $16$, the power consumption increases by $3.98\%$, as shown in Table~\ref{same-set-diff-way}.  This result indicates that \scheme\ design can work with different sizes of cache and various number of ways with reasonable overhead. 
As shown in Table~\ref{skipcnn-vs-baseline}, \scheme\ increases resource usage and the power consumption by $1.135\times$. The majority of this increment is related to the power of Signals and Logic which is the result of using adaptable cache-based structure in the proposed design. However, as explained in $\S$~\ref{sec:implement}, since we fixed the cache size and number of ways to an optimal value, the \scheme\ resource consumption will be almost the same even if we increase the size of input images and number of PEs. 
\vspace{-0.15cm}

%% file: conclusion.tex
\vspace{-0.15cm}
We proposed a novel scheme based on RPQ to exploit the similarity of computations during DNN training in a hardware accelerator. The proposed scheme, called \scheme, uses a cache (\scache) to store signatures of recent input vectors along with the computed results. 
If a new input vector's signature matches with an existing signature in the \scache, the already-computed result is reused for the new vector. \scheme\ is the {\em first} work that exploits computational similarity using RPQ for accelerating DNN training in hardware.
We present a detailed design, workflow, and implementation of \scheme for multiple layers and dataflows. This work opens up a new direction for speeding up hardware accelerators by optimizing computations for the deep learning process. 
Our experimental evaluation with twelve different deep learning models shows that \scheme\ speeds up the training by $1.97\times$ with an accuracy similar to the baseline system.

%% file: main.bbl
\begin{thebibliography}{10}
\providecommand{\url}[1]{#1}
\csname url@samestyle\endcsname
\providecommand{\newblock}{\relax}
\providecommand{\bibinfo}[2]{#2}
\providecommand{\BIBentrySTDinterwordspacing}{\spaceskip=0pt\relax}
\providecommand{\BIBentryALTinterwordstretchfactor}{4}
\providecommand{\BIBentryALTinterwordspacing}{\spaceskip=\fontdimen2\font plus
\BIBentryALTinterwordstretchfactor\fontdimen3\font minus
  \fontdimen4\font\relax}
\providecommand{\BIBforeignlanguage}[2]{{%
\expandafter\ifx\csname l@#1\endcsname\relax
\typeout{** WARNING: IEEEtranS.bst: No hyphenation pattern has been}%
\typeout{** loaded for the language `#1'. Using the pattern for}%
\typeout{** the default language instead.}%
\else
\language=\csname l@#1\endcsname
\fi
#2}}
\providecommand{\BIBdecl}{\relax}
\BIBdecl

\bibitem{snapea}
V.~Akhlaghi, A.~Yazdanbakhsh, K.~Samadi, R.~K. Gupta, and H.~Esmaeilzadeh,
  ``Snapea: Predictive early activation for reducing computation in deep
  convolutional neural networks,'' in \emph{2018 ACM/IEEE 45th Annual
  International Symposium on Computer Architecture (ISCA)}, 2018, pp. 662--673.

\bibitem{cnvlutin}
J.~{Albericio}, P.~{Judd}, T.~{Hetherington}, T.~{Aamodt}, N.~E. {Jerger}, and
  A.~{Moshovos}, ``Cnvlutin: Ineffectual-neuron-free deep neural network
  computing,'' in \emph{2016 ACM/IEEE 43rd Annual International Symposium on
  Computer Architecture (ISCA)}, 2016, pp. 1--13.

\bibitem{fused-cnn}
M.~Alwani, H.~Chen, M.~Ferdman, and P.~Milder, ``Fused-layer cnn
  accelerators,'' in \emph{2016 49th Annual IEEE/ACM International Symposium on
  Microarchitecture (MICRO)}, 2016, pp. 1--12.

\bibitem{bengio2007scaling}
Y.~Bengio, Y.~LeCun \emph{et~al.}, ``Scaling learning algorithms towards ai,''
  \emph{Large-scale kernel machines}, vol.~34, no.~5, pp. 1--41, 2007.

\bibitem{random-projection}
\BIBentryALTinterwordspacing
E.~Bingham and H.~Mannila, ``Random projection in dimensionality reduction:
  Applications to image and text data,'' in \emph{Proceedings of the Seventh
  ACM SIGKDD International Conference on Knowledge Discovery and Data Mining},
  ser. KDD '01.\hskip 1em plus 0.5em minus 0.4em\relax New York, NY, USA:
  Association for Computing Machinery, 2001, p. 245–250. [Online]. Available:
  \url{https://doi.org/10.1145/502512.502546}
\BIBentrySTDinterwordspacing

\bibitem{bloom1970space}
B.~H. Bloom, ``Space/time trade-offs in hash coding with allowable errors,''
  \emph{Communications of the ACM}, vol.~13, no.~7, pp. 422--426, 1970.

\bibitem{ceze2006bulk}
\BIBentryALTinterwordspacing
L.~Ceze, J.~Tuck, J.~Torrellas, and C.~Cascaval, ``Bulk disambiguation of
  speculative threads in multiprocessors,'' in \emph{Proceedings of the 33rd
  Annual International Symposium on Computer Architecture}, ser. ISCA
  '06.\hskip 1em plus 0.5em minus 0.4em\relax USA: IEEE Computer Society, 2006,
  p. 227–238. [Online]. Available: \url{https://doi.org/10.1109/ISCA.2006.13}
\BIBentrySTDinterwordspacing

\bibitem{rounding-rpq}
\BIBentryALTinterwordspacing
M.~S. Charikar, ``Similarity estimation techniques from rounding algorithms,''
  in \emph{Proceedings of the Thiry-Fourth Annual ACM Symposium on Theory of
  Computing}, ser. STOC '02.\hskip 1em plus 0.5em minus 0.4em\relax New York,
  NY, USA: Association for Computing Machinery, 2002, p. 380–388. [Online].
  Available: \url{https://doi.org/10.1145/509907.509965}
\BIBentrySTDinterwordspacing

\bibitem{diannao}
\BIBentryALTinterwordspacing
T.~Chen, Z.~Du, N.~Sun, J.~Wang, C.~Wu, Y.~Chen, and O.~Temam, ``Diannao: A
  small-footprint high-throughput accelerator for ubiquitous
  machine-learning,'' ser. ASPLOS '14.\hskip 1em plus 0.5em minus 0.4em\relax
  New York, NY, USA: Association for Computing Machinery, 2014, p. 269–284.
  [Online]. Available: \url{https://doi.org/10.1145/2541940.2541967}
\BIBentrySTDinterwordspacing

\bibitem{eyeriss}
Y.~{Chen}, J.~{Emer}, and V.~{Sze}, ``Eyeriss: A spatial architecture for
  energy-efficient dataflow for convolutional neural networks,'' in \emph{2016
  ACM/IEEE 43rd Annual International Symposium on Computer Architecture
  (ISCA)}, 2016, pp. 367--379.

\bibitem{dataflows}
Y.~Chen, J.~Emer, and V.~Sze, ``Using dataflow to optimize energy efficiency of
  deep neural network accelerators,'' \emph{IEEE Micro}, vol.~37, no.~3, pp.
  12--21, 2017.

\bibitem{dadiandao}
Y.~Chen, T.~Luo, S.~Liu, S.~Zhang, L.~He, J.~Wang, L.~Li, T.~Chen, Z.~Xu,
  N.~Sun, and O.~Temam, ``Dadiannao: A machine-learning supercomputer,'' in
  \emph{2014 47th Annual IEEE/ACM International Symposium on
  Microarchitecture}, 2014, pp. 609--622.

\bibitem{prime}
P.~Chi, S.~Li, C.~Xu, T.~Zhang, J.~Zhao, Y.~Liu, Y.~Wang, and Y.~Xie, ``Prime:
  A novel processing-in-memory architecture for neural network computation in
  reram-based main memory,'' in \emph{2016 ACM/IEEE 43rd Annual International
  Symposium on Computer Architecture (ISCA)}, 2016, pp. 27--39.

\bibitem{cicek-reuse}
N.~M. Cicek, L.~Ning, O.~Ozturk, and X.~Shen, ``General reuse-centric cnn
  accelerator,'' \emph{IEEE Transactions on Computers}, pp. 1--1, 2021.

\bibitem{cicek21}
\BIBentryALTinterwordspacing
N.~M. Cicek, X.~Shen, and O.~Ozturk, ``Energy efficient boosting of gemm
  accelerators for dnn via reuse,'' \emph{ACM Trans. Des. Autom. Electron.
  Syst.}, dec 2021, just Accepted. [Online]. Available:
  \url{https://doi.org/10.1145/3503469}
\BIBentrySTDinterwordspacing

\bibitem{coleman2019selection}
C.~Coleman, C.~Yeh, S.~Mussmann, B.~Mirzasoleiman, P.~Bailis, P.~Liang,
  J.~Leskovec, and M.~Zaharia, ``Selection via proxy: Efficient data selection
  for deep learning,'' \emph{arXiv preprint arXiv:1906.11829}, 2019.

\bibitem{binarynet}
M.~{Courbariaux} and Y.~{Bengio}, ``Binarynet: Training deep neural networks
  with weights and activations constrained to +1 or -1,'' in \emph{Conference
  on Neural Information Processing Systems (NIPS)}, 2016.

\bibitem{das-mixed}
D.~Das, N.~Mellempudi, D.~Mudigere, D.~Kalamkar, S.~Avancha, K.~Banerjee,
  S.~Sridharan, K.~Vaidyanathan, B.~Kaul, E.~Georganas, A.~Heinecke, P.~Dubey,
  J.~Corbal, N.~Shustrov, R.~Dubtsov, E.~Fomenko, and V.~Pirogov, ``Mixed
  precision training of convolutional neural networks using integer
  operations,'' in \emph{International Conference on Learning Representations
  (ICLR)}, 2018.

\bibitem{dist-large}
\BIBentryALTinterwordspacing
J.~Dean, G.~Corrado, R.~Monga, K.~Chen, M.~Devin, M.~Mao, M.~a. Ranzato,
  A.~Senior, P.~Tucker, K.~Yang, Q.~Le, and A.~Ng, ``Large scale distributed
  deep networks,'' in \emph{Advances in Neural Information Processing Systems},
  F.~Pereira, C.~Burges, L.~Bottou, and K.~Weinberger, Eds., vol.~25.\hskip 1em
  plus 0.5em minus 0.4em\relax Curran Associates, Inc., 2012. [Online].
  Available:
  \url{https://proceedings.neurips.cc/paper/2012/file/6aca97005c68f1206823815f66102863-Paper.pdf}
\BIBentrySTDinterwordspacing

\bibitem{bit-tactical}
\BIBentryALTinterwordspacing
A.~Delmas~Lascorz, P.~Judd, D.~M. Stuart, Z.~Poulos, M.~Mahmoud, S.~Sharify,
  M.~Nikolic, K.~Siu, and A.~Moshovos, ``Bit-tactical: A software/hardware
  approach to exploiting value and bit sparsity in neural networks,'' ser.
  ASPLOS '19.\hskip 1em plus 0.5em minus 0.4em\relax New York, NY, USA:
  Association for Computing Machinery, 2019, p. 749–763. [Online]. Available:
  \url{https://doi.org/10.1145/3297858.3304041}
\BIBentrySTDinterwordspacing

\bibitem{shidiannao}
Z.~Du, R.~Fasthuber, T.~Chen, P.~Ienne, L.~Li, T.~Luo, X.~Feng, Y.~Chen, and
  O.~Temam, ``Shidiannao: Shifting vision processing closer to the sensor,'' in
  \emph{2015 ACM/IEEE 42nd Annual International Symposium on Computer
  Architecture (ISCA)}, 2015, pp. 92--104.

\bibitem{jpeg-act}
\BIBentryALTinterwordspacing
R.~D. Evans, L.~Liu, and T.~M. Aamodt, ``Jpeg-act: Accelerating deep learning
  via transform-based lossy compression,'' in \emph{Proceedings of the ACM/IEEE
  47th Annual International Symposium on Computer Architecture}, ser. ISCA
  '20.\hskip 1em plus 0.5em minus 0.4em\relax IEEE Press, 2020, p. 860–873.
  [Online]. Available: \url{https://doi.org/10.1109/ISCA45697.2020.00075}
\BIBentrySTDinterwordspacing

\bibitem{Feldman2020}
\BIBentryALTinterwordspacing
D.~Feldman, \emph{Core-Sets: Updated Survey}.\hskip 1em plus 0.5em minus
  0.4em\relax Cham: Springer International Publishing, 2020, pp. 23--44.
  [Online]. Available: \url{https://doi.org/10.1007/978-3-030-29349-9_2}
\BIBentrySTDinterwordspacing

\bibitem{manyfold-rpq}
\BIBentryALTinterwordspacing
Y.~Freund, S.~Dasgupta, M.~Kabra, and N.~Verma, ``Learning the structure of
  manifolds using random projections,'' in \emph{Advances in Neural Information
  Processing Systems}, J.~Platt, D.~Koller, Y.~Singer, and S.~Roweis, Eds.,
  vol.~20.\hskip 1em plus 0.5em minus 0.4em\relax Curran Associates, Inc.,
  2007. [Online]. Available:
  \url{https://proceedings.neurips.cc/paper/2007/file/9fc3d7152ba9336a670e36d0ed79bc43-Paper.pdf}
\BIBentrySTDinterwordspacing

\bibitem{sparten}
\BIBentryALTinterwordspacing
A.~Gondimalla, N.~Chesnut, M.~Thottethodi, and T.~N. Vijaykumar, ``Sparten: A
  sparse tensor accelerator for convolutional neural networks,'' in
  \emph{Proceedings of the 52nd Annual IEEE/ACM International Symposium on
  Microarchitecture}, ser. MICRO '52.\hskip 1em plus 0.5em minus 0.4em\relax
  New York, NY, USA: Association for Computing Machinery, 2019, p. 151–165.
  [Online]. Available: \url{https://doi.org/10.1145/3352460.3358291}
\BIBentrySTDinterwordspacing

\bibitem{cnn-rev}
\BIBentryALTinterwordspacing
J.~Gu, Z.~Wang, J.~Kuen, L.~Ma, A.~Shahroudy, B.~Shuai, T.~Liu, X.~Wang,
  G.~Wang, J.~Cai, and T.~Chen, ``Recent advances in convolutional neural
  networks,'' \emph{Pattern Recogn.}, vol.~77, no.~C, p. 354–377, May 2018.
  [Online]. Available: \url{https://doi.org/10.1016/j.patcog.2017.10.013}
\BIBentrySTDinterwordspacing

\bibitem{dcompress}
S.~{Han}, H.~{Mao}, and W.~{Dally}, ``Deep compression: Compressing deep neural
  networks with pruning, trained quantization and huffman coding,'' in
  \emph{International Conference on Learning Representations (ICLR)}, 2016.

\bibitem{eie}
S.~Han, X.~Liu, H.~Mao, J.~Pu, A.~Pedram, M.~A. Horowitz, and W.~J. Dally,
  ``Eie: Efficient inference engine on compressed deep neural network,'' in
  \emph{2016 ACM/IEEE 43rd Annual International Symposium on Computer
  Architecture (ISCA)}, 2016, pp. 243--254.

\bibitem{ucnn}
K.~{Hegde}, J.~{Yu}, R.~{Agrawal}, M.~{Yan}, M.~{Pellauer}, and C.~{Fletcher},
  ``Ucnn: Exploiting computational reuse in deep neural networks via weight
  repetition,'' in \emph{2018 ACM/IEEE 45th Annual International Symposium on
  Computer Architecture (ISCA)}, 2018, pp. 674--687.

\bibitem{hennessy}
J.~L. Hennessy and D.~A. Patterson, \emph{Computer Architecture, Fifth Edition:
  A Quantitative Approach}, 5th~ed.\hskip 1em plus 0.5em minus 0.4em\relax San
  Francisco, CA, USA: Morgan Kaufmann Publishers Inc., 2011.

\bibitem{gist}
\BIBentryALTinterwordspacing
A.~Jain, A.~Phanishayee, J.~Mars, L.~Tang, and G.~Pekhimenko, ``Gist: Efficient
  data encoding for deep neural network training,'' in \emph{Proceedings of the
  45th Annual International Symposium on Computer Architecture}, ser. ISCA
  '18.\hskip 1em plus 0.5em minus 0.4em\relax IEEE Press, 2018, p. 776–789.
  [Online]. Available: \url{https://doi.org/10.1109/ISCA.2018.00070}
\BIBentrySTDinterwordspacing

\bibitem{tpu}
N.~P. Jouppi, C.~Young, N.~Patil, D.~Patterson, G.~Agrawal, R.~Bajwa, S.~Bates,
  S.~Bhatia, N.~Boden, A.~Borchers, R.~Boyle, P.-l. Cantin, C.~Chao, C.~Clark,
  J.~Coriell, M.~Daley, M.~Dau, J.~Dean, B.~Gelb, T.~V. Ghaemmaghami,
  R.~Gottipati, W.~Gulland, R.~Hagmann, C.~R. Ho, D.~Hogberg, J.~Hu, R.~Hundt,
  D.~Hurt, J.~Ibarz, A.~Jaffey, A.~Jaworski, A.~Kaplan, H.~Khaitan,
  D.~Killebrew, A.~Koch, N.~Kumar, S.~Lacy, J.~Laudon, J.~Law, D.~Le, C.~Leary,
  Z.~Liu, K.~Lucke, A.~Lundin, G.~MacKean, A.~Maggiore, M.~Mahony, K.~Miller,
  R.~Nagarajan, R.~Narayanaswami, R.~Ni, K.~Nix, T.~Norrie, M.~Omernick,
  N.~Penukonda, A.~Phelps, J.~Ross, M.~Ross, A.~Salek, E.~Samadiani, C.~Severn,
  G.~Sizikov, M.~Snelham, J.~Souter, D.~Steinberg, A.~Swing, M.~Tan,
  G.~Thorson, B.~Tian, H.~Toma, E.~Tuttle, V.~Vasudevan, R.~Walter, W.~Wang,
  E.~Wilcox, and D.~H. Yoon, ``In-datacenter performance analysis of a tensor
  processing unit,'' in \emph{2017 ACM/IEEE 44th Annual International Symposium
  on Computer Architecture (ISCA)}, 2017, pp. 1--12.

\bibitem{killamsetty2021grad}
K.~Killamsetty, S.~Durga, G.~Ramakrishnan, A.~De, and R.~Iyer, ``Grad-match:
  Gradient matching based data subset selection for efficient deep model
  training,'' in \emph{International Conference on Machine Learning}.\hskip 1em
  plus 0.5em minus 0.4em\relax PMLR, 2021, pp. 5464--5474.

\bibitem{killamsetty2021glister}
K.~Killamsetty, D.~Sivasubramanian, G.~Ramakrishnan, and R.~Iyer, ``Glister:
  Generalization based data subset selection for efficient and robust
  learning,'' in \emph{Proceedings of the AAAI Conference on Artificial
  Intelligence}, vol.~35, no.~9, 2021, pp. 8110--8118.

\bibitem{flexpoint}
U.~K\"{o}ster, T.~J. Webb, X.~Wang, M.~Nassar, A.~K. Bansal, W.~H. Constable,
  O.~H. Elibol, S.~Gray, S.~Hall, L.~Hornof, A.~Khosrowshahi, C.~Kloss, R.~J.
  Pai, and N.~Rao, ``Flexpoint: An adaptive numerical format for efficient
  training of deep neural networks,'' in \emph{Proceedings of the 31st
  International Conference on Neural Information Processing Systems}, ser.
  NIPS'17.\hskip 1em plus 0.5em minus 0.4em\relax Red Hook, NY, USA: Curran
  Associates Inc., 2017, p. 1740–1750.

\bibitem{kumar2010self}
M.~Kumar, B.~Packer, and D.~Koller, ``Self-paced learning for latent variable
  models,'' \emph{Advances in neural information processing systems}, vol.~23,
  2010.

\bibitem{maeri}
\BIBentryALTinterwordspacing
H.~Kwon, A.~Samajdar, and T.~Krishna, ``Maeri: Enabling flexible dataflow
  mapping over dnn accelerators via reconfigurable interconnects,''
  \emph{SIGPLAN Not.}, vol.~53, no.~2, p. 461–475, 2018. [Online]. Available:
  \url{https://doi.org/10.1145/3296957.3173176}
\BIBentrySTDinterwordspacing

\bibitem{cnnapp}
Z.~Li, F.~Liu, W.~Yang, S.~Peng, and J.~Zhou, ``A survey of convolutional
  neural networks: Analysis, applications, and prospects,'' \emph{IEEE
  Transactions on Neural Networks and Learning Systems}, pp. 1--21, 2021.

\bibitem{pudiannao}
\BIBentryALTinterwordspacing
D.~Liu, T.~Chen, S.~Liu, J.~Zhou, S.~Zhou, O.~Teman, X.~Feng, X.~Zhou, and
  Y.~Chen, ``Pudiannao: A polyvalent machine learning accelerator,''
  \emph{SIGPLAN Not.}, vol.~50, no.~4, p. 369–381, Mar. 2015. [Online].
  Available: \url{https://doi.org/10.1145/2775054.2694358}
\BIBentrySTDinterwordspacing

\bibitem{dsg}
L.~Liu, L.~Deng, X.~Hu, M.~Zhu, G.~Li, Y.~Ding, and Y.~Xie, ``Dynamic sparse
  graph for efficient deep learning,'' in \emph{7th International Conference on
  Learning Representations, {ICLR} 2019}, 2019.

\bibitem{tdash}
M.~{Mahmoud}, I.~{Edo}, A.~H. {Zadeh}, O.~{Mohamed Awad}, G.~{Pekhimenko},
  J.~{Albericio}, and A.~{Moshovos}, ``Tensordash: Exploiting sparsity to
  accelerate deep neural network training,'' in \emph{2020 53rd Annual IEEE/ACM
  International Symposium on Microarchitecture (MICRO)}, 2020, pp. 781--795.

\bibitem{diffy}
M.~Mahmoud, K.~Siu, and A.~Moshovos, ``Diffy: a déjà vu-free differential
  deep neural network accelerator,'' in \emph{2018 51st Annual IEEE/ACM
  International Symposium on Microarchitecture (MICRO)}, 2018, pp. 134--147.

\bibitem{dist-survey}
\BIBentryALTinterwordspacing
R.~Mayer and H.-A. Jacobsen, ``Scalable deep learning on distributed
  infrastructures: Challenges, techniques, and tools,'' \emph{ACM Comput.
  Surv.}, vol.~53, no.~1, feb 2020. [Online]. Available:
  \url{https://doi.org/10.1145/3363554}
\BIBentrySTDinterwordspacing

\bibitem{mirzasoleiman2020coresets}
B.~Mirzasoleiman, J.~Bilmes, and J.~Leskovec, ``Coresets for data-efficient
  training of machine learning models,'' in \emph{International Conference on
  Machine Learning}.\hskip 1em plus 0.5em minus 0.4em\relax PMLR, 2020, pp.
  6950--6960.

\bibitem{adaptivedeepreuse}
L.~Ning, H.~Guan, and X.~Shen, ``Adaptive deep reuse: Accelerating cnn training
  on the fly,'' in \emph{2019 IEEE 35th International Conference on Data
  Engineering (ICDE)}, 2019, pp. 1538--1549.

\bibitem{dreuse}
L.~Ning and X.~Shen, ``Deep reuse: streamline cnn inference on the fly via
  coarse-grained computation reuse,'' \emph{Proceedings of the ACM
  International Conference on Supercomputing}, 2019.

\bibitem{scnn}
A.~Parashar, M.~Rhu, A.~Mukkara, A.~Puglielli, R.~Venkatesan, B.~Khailany,
  J.~Emer, S.~W. Keckler, and W.~J. Dally, ``Scnn: An accelerator for
  compressed-sparse convolutional neural networks,'' in \emph{2017 ACM/IEEE
  44th Annual International Symposium on Computer Architecture (ISCA)}, 2017,
  pp. 27--40.

\bibitem{pytorch}
A.~Paszke, S.~Gross, F.~Massa, A.~Lerer, J.~Bradbury, G.~Chanan, T.~Killeen,
  Z.~Lin, N.~Gimelshein, L.~Antiga \emph{et~al.}, ``Pytorch: An imperative
  style, high-performance deep learning library,'' \emph{Advances in neural
  information processing systems}, vol.~32, 2019.

\bibitem{outputstationary}
M.~Peemen, A.~A. Setio, B.~Mesman, and H.~Corporaal, ``Memory-centric
  accelerator design for convolutional neural networks,'' in \emph{2013 IEEE
  31st International Conference on Computer Design (ICCD)}.\hskip 1em plus
  0.5em minus 0.4em\relax IEEE, 2013, pp. 13--19.

\bibitem{summerge}
R.~B. Prabhakar, S.~Kuhar, R.~Agrawal, C.~J. Hughes, and C.~W. Fletcher,
  ``Summerge: An efficient algorithm and implementation for weight
  repetition-aware dnn inference,'' in \emph{Proceedings of the ACM
  International Conference on Supercomputing}, ser. ICS '21.\hskip 1em plus
  0.5em minus 0.4em\relax New York, NY, USA: Association for Computing
  Machinery, 2021, p. 279–290.

\bibitem{soft-pipe}
B.~R. Rau and C.~D. Glaeser, ``Some scheduling techniques and an easily
  schedulable horizontal architecture for high performance scientific
  computing,'' in \emph{Proceedings of the 14th Annual Workshop on
  Microprogramming}, ser. MICRO 14.\hskip 1em plus 0.5em minus 0.4em\relax IEEE
  Press, 1981, p. 183–198.

\bibitem{riera-reuse}
M.~Riera, J.-M. Arnau, and A.~Gonzalez, ``Computation reuse in dnns by
  exploiting input similarity,'' in \emph{2018 ACM/IEEE 45th Annual
  International Symposium on Computer Architecture (ISCA)}, 2018, pp. 57--68.

\bibitem{jason21}
\BIBentryALTinterwordspacing
J.~Servais and E.~Atoofian, ``Adaptive computation reuse for energy-efficient
  training of deep neural networks,'' \emph{ACM Trans. Embed. Comput. Syst.},
  vol.~20, no.~6, oct 2021. [Online]. Available:
  \url{https://doi.org/10.1145/3487025}
\BIBentrySTDinterwordspacing

\bibitem{isaac}
A.~Shafiee, A.~Nag, N.~Muralimanohar, R.~Balasubramonian, J.~P. Strachan,
  M.~Hu, R.~S. Williams, and V.~Srikumar, ``Isaac: A convolutional neural
  network accelerator with in-situ analog arithmetic in crossbars,'' in
  \emph{2016 ACM/IEEE 43rd Annual International Symposium on Computer
  Architecture (ISCA)}, 2016, pp. 14--26.

\bibitem{vgg}
\BIBentryALTinterwordspacing
K.~Simonyan and A.~Zisserman, ``Very deep convolutional networks for
  large-scale image recognition,'' in \emph{3rd International Conference on
  Learning Representations, {ICLR} 2015, San Diego, CA, USA, May 7-9, 2015,
  Conference Track Proceedings}, Y.~Bengio and Y.~LeCun, Eds., 2015. [Online].
  Available: \url{http://arxiv.org/abs/1409.1556}
\BIBentrySTDinterwordspacing

\bibitem{uspe}
M.~Song, J.~Zhao, Y.~Hu, J.~Zhang, and T.~Li, ``Prediction based execution on
  deep neural networks,'' in \emph{2018 ACM/IEEE 45th Annual International
  Symposium on Computer Architecture (ISCA)}, 2018, pp. 752--763.

\bibitem{toneva2018empirical}
M.~Toneva, A.~Sordoni, R.~T.~d. Combes, A.~Trischler, Y.~Bengio, and G.~J.
  Gordon, ``An empirical study of example forgetting during deep neural network
  learning,'' \emph{arXiv preprint arXiv:1812.05159}, 2018.

\bibitem{transformer}
\BIBentryALTinterwordspacing
A.~Vaswani, N.~Shazeer, N.~Parmar, J.~Uszkoreit, L.~Jones, A.~N. Gomez,
  L.~Kaiser, and I.~Polosukhin, ``Attention is all you need,'' 2017. [Online].
  Available: \url{https://arxiv.org/abs/1706.03762}
\BIBentrySTDinterwordspacing

\bibitem{virtex}
\BIBentryALTinterwordspacing
Xilinx, ``Virtex 7 fpga.'' [Online]. Available:
  \url{https://www.xilinx.com/products/silicon-devices/fpga/virtex-7.html}
\BIBentrySTDinterwordspacing

\bibitem{vivado}
\BIBentryALTinterwordspacing
Xilinx, ``Vivado.'' [Online]. Available:
  \url{https://www.xilinx.com/products/design-tools/vivado.html}
\BIBentrySTDinterwordspacing

\bibitem{face-rec}
\BIBentryALTinterwordspacing
J.~Xing, G.~Fang, J.~Zhong, and J.~Li, ``Application of face recognition based
  on cnn in fatigue driving detection,'' in \emph{Proceedings of the 2019
  International Conference on Artificial Intelligence and Advanced
  Manufacturing}, ser. AIAM 2019.\hskip 1em plus 0.5em minus 0.4em\relax New
  York, NY, USA: Association for Computing Machinery, 2019. [Online].
  Available: \url{https://doi.org/10.1145/3358331.3358387}
\BIBentrySTDinterwordspacing

\bibitem{procrustes}
D.~Yang, A.~Ghasemazar, X.~Ren, M.~Golub, G.~Lemieux, and M.~Lis, ``Procrustes:
  a dataflow and accelerator for sparse deep neural network training,'' in
  \emph{2020 53rd Annual IEEE/ACM International Symposium on Microarchitecture
  (MICRO)}, 2020, pp. 711--724.

\bibitem{Fan2017Learning}
T.~Q. J. B. T.-Y.~L. Yang~Fan, Fei~Tian, ``Learning what data to learn,'' 2017.

\bibitem{eager-prune}
J.~Zhang, X.~Chen, M.~Song, and T.~Li, ``Eager pruning: Algorithm and
  architecture support for fast training of deep neural networks,'' in
  \emph{2019 ACM/IEEE 46th Annual International Symposium on Computer
  Architecture (ISCA)}, 2019, pp. 292--303.

\bibitem{echo}
\BIBentryALTinterwordspacing
B.~Zheng, N.~Vijaykumar, and G.~Pekhimenko, ``Echo: Compiler-based gpu memory
  footprint reduction for lstm rnn training,'' in \emph{Proceedings of the
  ACM/IEEE 47th Annual International Symposium on Computer Architecture}, ser.
  ISCA '20.\hskip 1em plus 0.5em minus 0.4em\relax IEEE Press, 2020, p.
  1089–1102. [Online]. Available:
  \url{https://doi.org/10.1109/ISCA45697.2020.00092}
\BIBentrySTDinterwordspacing

\bibitem{ttq}
C.~Zhu, S.~Han, H.~Mao, and W.~J. Dally, ``Trained ternary quantization,'' in
  \emph{5th International Conference on Learning Representations, {ICLR} 2017},
  2017.

\end{thebibliography}
